\documentclass[10pt]{article}

\pdfoutput=1

\usepackage[super,sort&compress,comma]{natbib} 
\usepackage{mhchem}
\usepackage{times,mathptmx}
\usepackage{sectsty}
\usepackage{balance} 
\usepackage{graphicx} 
\usepackage{lastpage}
\usepackage[format=plain,justification=raggedright,singlelinecheck=false,font=small,labelfont=bf,labelsep=space]{caption} 
\usepackage{fancyhdr}
\usepackage{amssymb,graphicx,lineno}
\usepackage{rotating}
\usepackage{enumerate}
\usepackage{setspace}
\usepackage{multirow, color}
\usepackage[bookmarks=false, pagebackref, colorlinks, allcolors=blue]{hyperref}
\usepackage{siunitx}
\usepackage{listings}
\usepackage{appendix}
\usepackage{setspace}
\usepackage{booktabs,subcaption,threeparttable}
\usepackage{amsmath}
\usepackage{hyperref}
\usepackage{mhchem}
\usepackage{array}
\usepackage{titlesec}

\newcommand{\beq}{\begin{equation}}
\newcommand{\beql}[1]{\begin{equation}\label{#1}}
\newcommand{\eeq}{\end{equation}}
\newcommand{\eeqp}{.\end{equation}}
\newcommand{\eeqc}{,\end{equation}}
\newcommand{\ie}{\textit{i.e.},~}
\newcolumntype{C}[1]{>{\centering\arraybackslash}p{#1}}
\makeatletter
\def\p@subsubsection{}
\def\p@subsection{}
\def\p@section{}
\makeatother

\begin{document}

\thispagestyle{plain}
\fancypagestyle{plain}{
\renewcommand{\headrulewidth}{1pt}}
\renewcommand{\thefootnote}{\fnsymbol{footnote}}
\renewcommand\footnoterule{\vspace*{1pt}%
\hrule width 3.4in height 0.4pt \vspace*{5pt}} 

\makeatletter 
\def\subsubsection{\@startsection{subsubsection}{3}{10pt}{-1.25ex plus -1ex minus -.1ex}{0ex plus 0ex}{\normalsize\bf}} 
\def\paragraph{\@startsection{paragraph}{4}{10pt}{-1.25ex plus -1ex minus -.1ex}{0ex plus 0ex}{\normalsize\textit}} 
\renewcommand\@biblabel[1]{#1}            
\renewcommand\@makefntext[1]%
{\noindent\makebox[0pt][r]{\@thefnmark\,}#1}
\makeatother 
\renewcommand{\figurename}{\small{Fig.}~}
\sectionfont{\large}
\subsectionfont{\normalsize} 

\fancyfoot{}
\fancyfoot[L]{\vspace{-7pt}\footnotesize{\sffamily{\href{http://arxiv.org/abs/1305.2759}{arXiv: 1305.2759}}}}
\fancyfoot[R]{\footnotesize{\sffamily{Nouranian et al.~(2014), 1--\pageref{LastPage} ~\textbar  \hspace{2pt}\thepage}}}
\fancyfoot[C]{\footnotesize{\sffamily{physics.chem-ph }}}
\fancyhead{}
\fancyhead[R]{\footnotesize{\sffamily{An interatomic potential for saturated hydrocarbons}}}
\renewcommand{\headrulewidth}{1pt} 
\renewcommand{\footrulewidth}{1pt}
\setlength{\arrayrulewidth}{1pt}
\setlength{\columnsep}{6.5mm}
\setlength\bibsep{1pt}

\noindent\LARGE{\textbf{An interatomic potential for saturated hydrocarbons based on the \\ modified embedded-atom method}}
\vspace{0.5cm}

\noindent\large{\textbf{S. Nouranian,\textit{$^{a}$} M.A. Tschopp,\textit{$^{a,b}$} S.R. Gwaltney,\textit{$^{c}$} M.I. Baskes,\textit{$^{d}$}$^{\ast}$ and M.F. Horstemeyer\textit{$^{a,e}$}}}
\vspace{0.5cm}

\noindent \normalsize{In this work, we developed an interatomic potential for saturated hydrocarbons using the modified embedded-atom method (MEAM), a reactive semi-empirical many-body potential based on density functional theory and pair potentials.  We parameterized the potential by fitting to a large experimental and first-principles (FP) database consisting of 1) bond distances, bond angles, and atomization energies at 0 K of a homologous series of alkanes and their select isomers from methane to $n$-octane, 2) the potential energy curves of H$_2$, CH, and C$_2$ diatomics, 3) the potential energy curves of hydrogen, methane, ethane, and propane dimers, i.e., (H$_2$)$_2$, (CH$_4$)$_2$, (C$_2$H$_6$)$_2$, and (C$_3$H$_8$)$_2$, respectively, and 5) pressure-volume-temperature ($PVT$) data of a dense high-pressure methane system with the density of 0.5534 g/cc.  We compared the atomization energies and geometries of a range of linear alkanes, cycloalkanes, and free radicals calculated from the MEAM potential to those calculated by other commonly used reactive potentials for hydrocarbons, i.e., second-generation reactive empirical bond order (REBO) and reactive force field (ReaxFF).  MEAM reproduced the experimental and/or FP data with accuracy comparable to or better than REBO or ReaxFF.  The experimental $PVT$ data for a relatively large series of methane, ethane, propane, and butane systems with different densities were predicted reasonably well by the MEAM potential.  Although the MEAM formalism has been applied to atomic systems with predominantly metallic bonding in the past, the current work demonstrates the promising extension of the MEAM potential to covalently bonded molecular systems, specifically saturated hydrocarbons and saturated hydrocarbon-based polymers.  The MEAM potential has already been parameterized for a large number of metallic unary, binary, ternary, carbide, nitride, and hydride systems, and extending it to saturated hydrocarbons provides a reliable and transferable potential for atomistic/molecular studies of complex material phenomena involving hydrocarbon-metal or polymer-metal interfaces, polymer-metal nanocomposites, fracture and failure in hydrocarbon-based polymers, etc.  The latter is especially true since MEAM is a reactive potential that allows for dynamic bond formation and bond breaking during simulation.  Our results show that MEAM predicts the energetics of two major chemical reactions for saturated hydrocarbons, i.e., breaking a C--C and a C--H bond, reasonably well.  However, the current parameterization does not accurately reproduce the energetics and structures of unsaturated hydrocarbons and, therefore, should not be applied to such systems.
}
\vspace{0.5cm}

\footnotetext{\textit{$^{\ast}$~Corresponding author, email: baskes@bagley.msstate.edu}}
\footnotetext{\textit{$^{a}$~Center for Advanced Vehicular Systems (CAVS), Mississippi State University, Mississippi State, MS 39762, USA}}
\footnotetext{\textit{$^{b}$~Engility Corporation, (on site at) U.S.~Army Research Laboratory, Aberdeen Proving Ground, MD 21005, USA }}
\footnotetext{\textit{$^{c}$~Department of Chemistry, Mississippi State University, Mississippi State, MS 39762, USA }}
\footnotetext{\textit{$^{d}$~Department of Aerospace Engineering, Mississippi State University, Mississippi State, MS 39762, USA }}
\footnotetext{\textit{$^{e}$~Department of Mechanical Engineering, Mississippi State University, Mississippi State, MS 39762, USA}}

\section{Introduction \label{sec:sec1}}

\singlespace

\indent The embedded-atom method (EAM), developed by Daw and Baskes \cite{Daw1983,Daw1984} in the early 1980s, is a semi-empirical $N$-body potential useful for the atomistic simulations of metal systems.  It has successfully been utilized to calculate the energetics and structures of complex metallic systems involving free surfaces, defects, grain boundaries, etc. \cite{Daw1993}  The potential was later modified by Baskes \cite{Bas1987,Bas1992} to include the directionality of bonding in covalent materials such as silicon and germanium, \cite{Bas1989} leading to the modified embedded-atom method (MEAM) \cite{Bas1992} introduced in 1992.  It has undergone several modifications and enhancements since then to include, for example, second nearest-neighbor interactions \cite{Lee2000,Lee2001,Lee2003} and, more recently, a multi-state formalism.\cite{Bas2007}  The unique feature of the MEAM formalism is its ability to reproduce the physical properties of a large number of fcc, \cite{Lee2003,Zha2004} bcc, \cite{Lee2001,Zha2003a} hcp, \cite{Bas1994,Hu2001} and diamond cubic \cite{Zha2003} crystal structures in unary, binary, ternary, and higher order \cite{Jel2012} metal systems with the same semi-empirical formalism.  MEAM is also both reliable and transferable \cite{Kim2009} in the sense that it accurately reproduces the physical properties of the element or alloy (reliability) and performs reasonably well under circumstances other than the ones used for its parameterization (transferability). \cite{Kim2009}  Horstemeyer \cite{Hor2012} has an excellent review of the MEAM potential in the context of a multi-scale modeling methodology (integrated computational materials engineering) for metals.

The MEAM formalism has traditionally been used for pure metals and impurities, binary and ternary alloys, and hydride, carbide, and nitride metal systems with great success. \cite{Lee2010}  In addition, complex nanostructured systems have been studied using various MEAM-based potentials.  For example, Xiao et~al.~\cite{Xia2009} calculated the interaction of carbon nanotubes with nickel (Ni) nanoparticles, and Uddin et~al.~\cite{Udd2010} recently studied the mechanical properties of carbon nanotube-Ni composites using the MEAM potential.  We extend the MEAM formalism in the current paper to saturated hydrocarbons with the ultimate aim of capturing the energetics and geometries of commercially important hydrocarbon-based polymers (polyolefins) such as polyethylene and polypropylene.  Potentials such as MM3, \cite{All1989,Lii1989,Lii1989a} MM4, \cite{All1996} DREIDING, \cite{May1990} first- \cite{Bre1990} and second-generation reactive empirical bond order (REBO), \cite{Bre2002} reactive force field (ReaxFF), \cite{Van2001} charge-optimized many-body (COMB) potential, \cite{Yu2007,Shan2010} and condensed-phase optimized molecular potentials for atomistic simulation studies (COMPASS) \cite{Sun1998} have been used for hydrocarbon simulations, but of these potentials, only REBO, ReaxFF, and COMB are reactive and can allow for bond breaking.  Furthermore, most of these potentials are not suitable for hydrocarbon-metal systems, with only ReaxFF \cite{Somers2013, Castro2013, Monti2013, Kim2013} and COMB \cite{Liang2012} having been used in the past to study hydrocarbon-metal interactions.  Liang et~al.~\cite{Liang2013} have recently reviewed the use of reactive potentials for advanced atomistic simulations.

In this paper, we develop a new set of parameters within the MEAM framework to describe the interactions and equilibrium geometries of saturated hydrocarbons, specifically bond distances, bond angles, and atomization energies at 0 K.  We show that MEAM gives a comparable or more accurate reproduction of these properties relative to experimental and first-principles (FP) data in comparison with REBO and ReaxFF.  We also reproduce the potential energy curves of H$_2$, CH, and C$_2$ diatomics and (H$_2$)$_2$, (CH$_4$)$_2$, (C$_2$H$_6$)$_2$, and (C$_3$H$_8$)$_2$ dimer configurations and predict the pressure-volume-temperature ($PVT$) relationships of a series of select methane, ethane, propane, and butane systems in a reasonable agreement with the experimental data.  The energetics of C--H and C--C bond breaking in methane and ethane and the heat of reaction for select chemical reactions are also presented.  MEAM gives reasonable predictions of the energies associated with these two major chemical reactions in saturated hydrocarbons. The development of the first MEAM-based interatomic potential for saturated hydrocarbons and hydrocarbon-based polymers is a step towards reliably simulating systems and phenomena that have hitherto been difficult to study, such as reactive multicomponent (organics/metal) systems, polymer-metal interfaces and nanocomposites, fracture and crack growth in polymers, etc.

This paper is organized in the following manner.  In Section \ref{sec:sec2} the theory of the MEAM formalism is reviewed.  In Section \ref{sec:sec3}, the potential development and parameterization is described.  The results are given in Section \ref{sec:sec4}.

\section{Theory \label{sec:sec2}}

In the EAM and MEAM formalisms \cite{Daw1983,Daw1984,Bas1992} the total energy of a system of atoms ($E_{tot}$) is given by

\beql{eq:eq1}
E_{tot}=\sum_i\left[F_{\tau_i}\left(\bar{\rho}_i\right)+\frac{1}{2}\sum_{j(\neq{i})}S_{ij}\phi_{\tau_i\tau_j}\left(R_{ij}\right)\right]
\eeqc

\noindent where $F_{\tau_i}$ is the embedding energy function for element type $\tau_i$, which is defined as the energy required to embed an atom of element type $\tau_i$ in the background electron density $\bar{\rho}_i$ at site $i$, $S_{ij}$ is the screening factor between atoms at sites $i$ and $j$ (defined in Eqs.~\ref{eq:eq21} and \ref{eq:eq24}), and $\phi_{\tau_i\tau_j}$ is the pair interaction between atoms of element types $\tau_i$ and $\tau_j$ at sites $i$ and $j$ at the separation distance of $R_{ij}$.  To emphasize the multi-component nature of the model, the element type of the atom at site $i$ is denoted as $\tau_i$ in this manuscript to distinguish it from site designation $i$, and the screening factor is explicitly separated from the pair potential.  The embedding function is given by the specific simple form 
 
\beql{eq:eq2}
F_{\tau_i}\left(\bar{\rho}_i\right)=\left\{ 
  \begin{array}{l l}
    A_{\tau_i}{E}_{\tau_i}^0\dfrac{\bar{\rho}_i}{\bar{\rho}_{\tau_i}^0}\left(\textrm{ln}\dfrac{\bar{\rho}_i}{\bar{\rho}_{\tau_i}^0}\right) & \quad \textrm{if } \bar{\rho}_i \geq 0\\
    -A_{\tau_i}{E}_{\tau_i}^0\dfrac{\bar{\rho}_i}{\bar{\rho}_{\tau_i}^0} & \quad \textrm{if } \bar{\rho}_i < 0\\
  \end{array} \right.
\eeqc

\noindent where $A_{\tau_i}$ is a scaling factor, $E_{\tau_i}^0$ is the sublimation (cohesive) energy, and $\bar{\rho}_{\tau_i}^0$ is the background electron density for the reference structure of the atom of element type ${\tau_i}$ at site $i$.  For most elements, the reference structure is the equilibrium structure of the element in its reference state.  However, the reference structure of carbon is taken as diamond.  We will denote the properties of the equilibrium reference state with a superscript zero.  The analytic continuation of the embedding function for negative electron densities was considered as a computational convenience to prevent systems from entering this unphysical regime.  The origin of negative electron densities arises below in Eq.~\ref{eq:eq5}.  The MEAM formalism introduces directionality in bonding between atoms through decomposing $\bar{\rho}_i$ into spherically symmetric ($\rho_i^{(0)}$) and angular ($\rho_i^{(1)}$, $\rho_i^{(2)}$, and $\rho_i^{(3)}$) partial electron densities \cite{Bas1992,Lee2010,Val2006} as given by

\beql{eq:eq3a}
\rho_i^{(0)}=\sum_{j\neq{i}}S_{ij}\rho_{\tau_i}^{a(0)}\left(R_{ij}\right)
\eeqc

\beql{eq:eq3b}
\left(\rho_i^{(1)}\right)^2=\dfrac{\sum\limits_{\alpha} \left[\sum\limits_{j\neq{i}} \dfrac{R_{ij}^\alpha}{R_{ij}}S_{ij}t_{\tau_j}^{(1)}\rho_{\tau_j}^{a(1)}\left(R_{ij}\right)\right]^2\rho_i^{(0)}}{\sum\limits_{j\neq{i}} S_{ij}\left(t_{\tau_j}^{(1)}\right)^2\rho_{\tau_j}^{a(0)}\left(R_{ij}\right)}
\eeqc

\beql{eq:eq3c}
\left(\rho_i^{(2)}\right)^2=\dfrac{\left\{\sum\limits_{\alpha,\beta} \left[\sum\limits_{j\neq{i}} \dfrac{R_{ij}^\alpha{R}_{ij}^\beta}{R_{ij}^2}S_{ij}t_{\tau_j}^{(2)}\rho_{\tau_j}^{a(2)}\left(R_{ij}\right)\right]^2-\dfrac{1}{3}\left[\sum\limits_{j\neq{i}} S_{ij}t_{\tau_j}^{(2)}\rho_{\tau_j}^{a(2)}\left(R_{ij}\right)\right]^2\right\}\rho_i^{(0)}}{\sum\limits_{j\neq{i}} S_{ij}\left(t_{\tau_j}^{(2)}\right)^2\rho_{\tau_j}^{a(0)}\left(R_{ij}\right)}
\eeqc

\beql{eq:eq3d}
\left(\rho_i^{(3)}\right)^2=\dfrac{\left\{\sum\limits_{\alpha,\beta,\gamma} \left[\sum\limits_{j\neq{i}} \dfrac{R_{ij}^\alpha{R}_{ij}^\beta{R}_{ij}^\gamma}{R_{ij}^3}S_{ij}t_{\tau_j}^{(3)}\rho_{\tau_j}^{a(3)}\left(R_{ij}\right)\right]^2-\dfrac{3}{5}\sum\limits_{\alpha}\left[\sum\limits_{j\neq{i}} \dfrac{R_{ij}^\alpha}{R_{ij}}S_{ij}t_{\tau_j}^{(3)}\rho_{\tau_j}^{a(3)}\left(R_{ij}\right)\right]^2\right\}\rho_i^{(0)}}{\sum\limits_{j\neq{i}} S_{ij}\left(t_{\tau_j}^{(3)}\right)^2\rho_{\tau_j}^{a(0)}\left(R_{ij}\right)}
\eeqp

$\rho_{\tau_j}^{a(h)}$ ($h=\left\{0,1,2,3\right\}$) indicate the atomic electron densities from atom of element type ${\tau_j}$  at site $j$  at distance $R_{ij}$ from site $i$.  $R_{ij}^\alpha$, $R_{ij}^\beta$, and $R_{ij}^\gamma$  represent the $\alpha$, $\beta$, and $\gamma$ components of the distance vector between atoms at sites $i$ and $j$, respectively, and $t_{\tau_j}^{(h)}$  ($h=\left\{1,2,3\right\}$) are adjustable element-dependent parameters.  The equivalence between these expressions and an expansion in Legendre polynomials has been discussed previously. \cite{Bas1992}  As above, we have carefully denoted the element types of the atoms, and separated the screening from the atomic electron densities.   Note that Eq.~\ref{eq:eq3a} is the simple linear superposition of atomic densities of the EAM formalism, \cite{Daw1983,Daw1984} and Eqs.~\ref{eq:eq3b}-\ref{eq:eq3d} reduce to more familiar forms in the original MEAM paper by Baskes \cite{Bas1992} for a single-component system.  The above partial electron densities can be combined in different ways to give the total background electron density at site $i$ ($\bar{\rho}_i$).  Here, we adopt one of the most widely used forms, \cite{Jel2012,Kim2009,Bas1999} which is given by

\beql{eq:eq4}
\bar{\rho}_i=\rho_i^{(0)}G\left(\Gamma_i\right)
\eeqc

\beql{eq:eq5}
G\left(\Gamma_i\right)=\left\{ 
  \begin{array}{l l}
    \sqrt{1+\Gamma_i} & \quad \textrm{if } \Gamma_i \geq -1\\
    -\sqrt{\left|1+\Gamma_i\right|} & \quad \textrm{if } \Gamma_i < -1\\
  \end{array} \right.
\eeqc

\beql{eq:eq6}
\Gamma_i=\sum\limits_{h=1}^3\bar{t}_i^{(h)}\left[\dfrac{\rho_i^{(h)}}{\rho_i^{(0)}}\right]^2
\eeqc

\beql{eq:eq7}
\bar{t}_i^{(h)}=\dfrac{1}{\rho_i^{(0)}}\sum\limits_{j\neq{i}} t_{\tau_j}^{(h)}\rho_{\tau_j}^{a(0)}S_{ij}
\eeqp

In the absence of angular contributions to the density, $\Gamma_i=0$, $G(\Gamma_i)=1$, and the model reduces to the EAM formalism.  For systems with negative $t_{\tau_j}^{(h)}$ values in certain geometries, $\Gamma_i < -1$, and for computational convenience we perform an analytic continuation of $G\left(\Gamma_i\right)$.  We choose to do this by allowing $G\left(\Gamma_i\right)$ and, hence, $\bar{\rho}_i$ to become less than zero. 

If we apply Eqs.~\ref{eq:eq4} and \ref{eq:eq6} to the equilibrium reference structure, we obtain

\beql{eq:eq8}
\bar{\rho}_\tau^0=Z_\tau^0\rho_\tau^0G\left(\Gamma_\tau^0\right)
\eeqc

\beql{eq:eq9}
\Gamma_\tau^0=\sum\limits_{h=1}^3 t_\tau^{(h)}s_\tau^{(h)}\left(\dfrac{1}{Z_\tau^0}\right)^2
\eeqc

\noindent where we have assumed that the reference structure has only first nearest-neighbor interactions.  In Eq.~\ref{eq:eq8}, $\rho_\tau^0$ is an element-dependent electron density scaling factor, and $Z_\tau^0$ is the first nearest-neighbor coordination number of the reference structure. $s_\tau^{(h)}$ ($h=\left\{1,2,3\right\}$) are ``shape factors'' that depend on the reference structure for element type $\tau$.  The shape factors are given in the original MEAM paper by Baskes. \cite{Bas1992}  The atomic electron density for element type $\tau$ is calculated from

\beql{eq:eq10}
\rho_\tau^{a(h)}\left(R\right)=\rho_\tau^0e^{-\beta_\tau^{(h)}\left(\dfrac{R}{R_\tau^0}-1\right)}
\eeqc

\noindent where $\beta_\tau^{(h)}$ ($h=\left\{0,1,2,3\right\}$) are adjustable element-dependent parameters, and $R_\tau^0$ is the nearest-neighbor distance in the equilibrium reference structure for the element type $\tau$. 

The pair interaction for like atoms of element type $\tau$ can be calculated using a first nearest-neighbor (1NN) \cite{Bas1992} or second nearest-neighbor (2NN) \cite{Lee2000,Lee2010} formalism.  In this work, the former is used \cite{Jel2012,Kim2009} and is given by

\beql{eq:eq11}
\phi_{\tau\tau}\left(R\right)=\dfrac{2}{Z_\tau^0}\left\{E_\tau^u\left(R\right)-F_\tau\left[\bar{\rho}_\tau^{ref}\left(R\right)\right]\right\}
\eeqp

In this equation $\bar{\rho}_\tau^{ref}\left(R\right)$ is the background electron density in the reference structure evaluated from Eqs.~\ref{eq:eq4}-\ref{eq:eq7} at a nearest-neighbor distance of $R$ and is given by

\beql{eq:eq12}
\bar{\rho}_\tau^{ref}\left(R\right)=Z_\tau^0\rho_\tau^0G\left(\Gamma_\tau^{ref}\right)
\eeqc

\beql{eq:eq13}
\Gamma_\tau^{ref}=\sum\limits_{h=1}^3 t_\tau^{(h)}s_\tau^{(h)}\left(\dfrac{\rho_\tau^{a(h)}}{Z_\tau^0\rho_\tau^{a(0)}}\right)^2
\eeqc

\noindent and $E_\tau^u$ is the universal equation of state (UEOS) of Rose et~al.~\cite{Ros1984} for element type $\tau$ given by

\beql{eq:eq14}
E_\tau^u\left(R\right)=-E_\tau^0\left[1+a^\ast+\delta\dfrac{R_\tau^0}{R}\left(a^\ast\right)^3\right]e^{-a^\ast}
\eeqc

\beql{eq:eq15}
a^\ast=\alpha_\tau^0\left(\dfrac{R}{R_\tau^0}-1\right)
\eeqc

\beql{eq:eq16}
\delta = \left\{ 
  \begin{array}{l l}
    \delta_\tau^a & \quad \textrm{if } a^\ast \geq 0\\
    \delta_\tau^r & \quad \textrm{if } a^\ast < 0\\
  \end{array} \right.
\eeqc

\beql{eq:eq17a}
\alpha_\tau^0=\sqrt{\dfrac{9K_\tau^0\Omega_\tau^0}{E_\tau^0}}
\eeqc

or

\beql{eq:eq17b}
\alpha_\tau^0=\sqrt{\dfrac{k_\tau^0}{E_\tau^0}}R_\tau^0
\eeqp

In the above equations $K_\tau^0$ ($k_\tau^0$) and $\Omega_\tau^0$ are the bulk modulus (spring constant) and the atomic volume of the reference structure, respectively, and $\delta$ is an adjustable, element-dependent parameter that has two components, attractive $\delta_\tau^a$ and repulsive $\delta_\tau^r$.  Eq.~\ref{eq:eq17a} is used when the reference structure is a three-dimensional (3D) crystal and Eq.~\ref{eq:eq17b} is used when the reference structure is a diatomic.

The pair interaction for unlike atoms of element types $\tau$ and $\upsilon$ is similarly obtained from the reference structure of the unlike atoms.  For this work, the reference structure is taken as the heteronuclear diatomic, which gives

\beql{eq:eq18}
\phi_{\tau\upsilon}\left(R\right)=\dfrac{1}{Z_{\tau\upsilon}^0}\left\{2E_{\tau\upsilon}^u\left(R\right)-F_\tau\left[\bar{\rho}_\upsilon^{d}\left(R\right)\right]-F_\upsilon\left[\bar{\rho}_\tau^{d}\left(R\right)\right]\right\}
\eeqc

\noindent where $Z_{\tau\upsilon}^0=1$ is the coordination number for the diatomic and

\beql{eq:eq19}
\bar{\rho}_\tau^{d}\left(R\right)=\rho_\tau^{a(0)}G\left(\Gamma_\tau^d\right)
\eeqc

\beql{eq:eq20}
\Gamma_\tau^d=\sum\limits_{h=1}^3 t_\tau^{(h)}s_d^{(h)}\left(\dfrac{\rho_\tau^{a(h)}}{\rho_\tau^{a(0)}}\right)^2
\eeqc

\noindent where the shape factors $s_d^{(h)}$ are those for a diatomic.  The UEOS $E_{\tau\upsilon}^u$ is given by Eqs.~\ref{eq:eq14}-\ref{eq:eq17b} using parameters $E_{\tau\upsilon}^0$, $R_{\tau\upsilon}^0$, $k_{\tau\upsilon}^0$, $\delta_{\tau\upsilon}^a$, and $\delta_{\tau\upsilon}^r$.

The screening factor $S_{ij}$ is defined as the product of all screening factors $S_{ikj}$, where the interaction between atoms at sites $i$ and $j$ are screened by neighboring atoms at site $k$ as given by

\beql{eq:eq21}
S_{ij}=\prod\limits_{k\neq{i,j}}S_{ikj}
\eeqp

If it is assumed that all three sites $i$, $j$, and $k$ lie on an ellipse on the $xy$-plane with sites $i$ and $j$ on the x-axis, the following equation can be derived:

\beql{eq:eq22}
x^2+\dfrac{1}{C}y^2=\left(\dfrac{1}{2}R_{ij}\right)^2
\eeqc

\noindent where

\beql{eq:eq23}
C_{ikj}=\dfrac{2\left(X_{ik}+X_{kj}\right)-\left(X_{ik}-X_{kj}\right)^2-1}{1-\left(X_{ik}-X_{kj}\right)^2}
\eeqp

In the above equation $X_{ik}=\left(R_{ik}/R_{ij}\right)^2$ and $X_{kj}=\left(R_{kj}/R_{ij}\right)^2$.  The screening factor $S_{ikj}$ for like atoms is defined as

\beql{eq:eq24}
S_{ikj}=f_c\left(\dfrac{C_{ikj}-C_{min}\left(\tau_i,\tau_k,\tau_j\right)}{C_{max}\left(\tau_i,\tau_k,\tau_j\right)-C_{min}\left(\tau_i,\tau_k,\tau_j\right)}\right)
\eeqc

\noindent where $C_{min}\left(\tau_i,\tau_k,\tau_j\right)$ and $C_{max}\left(\tau_i,\tau_k,\tau_j\right)$  determine the extent of screening of atoms of element type $\tau$ at sites $i$ and $j$ by an atom at site $k$.  Similar expressions can be written for the screening of unlike atoms.  The smooth cutoff function $f_c$ is defined as

\beql{eq:eq25}
f_c\left(x\right)=\left\{ 
  \begin{array}{l l}
    1 & \quad \textrm{if } x \geq 1\\
    \left[1-\left(1-x\right)^4\right]^2 & \quad \textrm{if } 0<x<1\\
    0 & \quad \textrm{if } x \leq 0\\
  \end{array} \right.
\eeqp

\noindent $S_{ij}=1$  means that the interaction between atoms at sites $i$ and $j$ is not screened, while $S_{ij}=0$ means the interaction is completely screened.

\section{Potential Parameterization \label{sec:sec3}}

The MEAM formalism presented in Eqs.~\ref{eq:eq1}-\ref{eq:eq25} requires 16 independent model parameters for each element type $\tau$, \ie $E_\tau^0$, $R_\tau^0$, $\alpha_\tau^0$, $\delta_\tau^a$, and $\delta_\tau^r$ for the universal equation of state (Eq.~\ref{eq:eq14}); $\beta_\tau^{(0)}$, $\beta_\tau^{(1)}$, $\beta_\tau^{(2)}$, $\beta_\tau^{(3)}$, $t_\tau^{(1)}$, $t_\tau^{(2)}$, $t_\tau^{(3)}$, and $\rho_\tau^0$ for the electron densities (Eqs.~\ref{eq:eq3a}-\ref{eq:eq10}); $A_\tau$ for the embedding function $F_\tau$ (Eq.~\ref{eq:eq2}); and $C_{min}$  and $C_{max}$ for the screening factor (Eqs.~\ref{eq:eq21}-\ref{eq:eq25}).  In the current MEAM formalism for a single element, the model is independent of $\rho_\tau^0$; hence, $\rho_\tau^0=1$  is taken for one of the elements.  For a diatomic composed of elements $\tau$ and $\upsilon$, 13 additional independent parameters are required, \ie $E_{\tau\upsilon}^0$, $R_{\tau\upsilon}^0$, $\alpha_{\tau\upsilon}^0$, $\delta_{\tau\upsilon}^a$, and $\delta_{\tau\upsilon}^r$, four $C_{min}$, and four $C_{max}$ values. 

In this work, we parameterized the elements carbon and hydrogen and the diatomic CH with the reference structures of diamond for carbon ($Z_C^0=4$), diatomic H$_2$ for hydrogen ($Z_H^0=1$), and diatomic CH for hydrocarbons ($Z_{CH}^0=1$).  The choice of reference structure for hydrocarbons is not unique.  For example, Valone et~al.~\cite{Val2006a} used ethylene as a reference structure in their work.  As initial starting parameters for this potential, we utilized the MEAM parameters for carbon from the work of Xiao et~al.~\cite{Xia2009} and the parameters for hydrogen from the work of Baskes. \cite{Bas1992}  The fitting procedure for the pure element parameters is described in detail by Baskes. \cite{Bas1992}  For the potential parameterization, we utilized a parameter fitting database consisting of 1) experimental bond distances, \cite{Lid2009} bond angles, \cite{Lid2009} and atomization energies at 0 K of a homologous series of alkanes and their isomers from methane to $n$-octane \cite{Nis2011,Kar2009} 2) potential energy curves of H$_2$, C$_2$, and CH diatomics, generated in this work from FP calculations, 3) FP interaction energy curves of (H$_2$)$_2$ (H$_2$ dimer), \cite{Bur1982} (CH$_4$)$_2$ (methane dimer), \cite{Szc1990} (C$_2$H$_6$)$_2$ (ethane dimer), \cite{Row2001} and (C$_3$H$_8$)$_2$ (propane dimer) \cite{Jal2002} in select molecular orientations, and 4) the pressure-volume-temperature ($PVT$) experimental data for a dense methane system. \cite{Rob1969}  This specific database was selected to provide ample experimental and/or FP data related to the energetics and structures (geometries) of representative alkane molecules as well as the intermolecular interactions between the molecules in a real hydrocarbon system for the purpose of potential parameterization.  This is also referred to as calibration of the model. The FP data for H$_2$, C$_2$, and CH diatomics were generated with the CCSD(2) \cite{Gwa2001} ab initio method and the aug-cc-pVTZ basis set \cite{Ken1992} using Q-Chem\textregistered~ quantum chemistry software (V3.2) \cite{Sha2006} and restricted core orbitals.  CCSD(2) is a high-accuracy FP method designed to calculate bond breaking with great precision. \cite{Gwa2003}  All molecules are in their ground electronic states.

The MEAM fitting procedure involved a stepwise and iterative effort to first capture the energy versus distance characteristics of the H$_2$, CH, and C$_2$ diatomics.  Next, we fit the atomization energies of the linear alkanes to the experimental data that were first corrected for the zero-point energy (ZPE).  The bond distances and bond angles of the first four alkanes and one butane isomer were then fit to the experimental data.  To enable a reliable prediction of the intermolecular forces, we further fit the MEAM parameters to the interaction energy curves of (H$_2$)$_2$, (CH$_4$)$_2$, (C$_2$H$_6$)$_2$, and (C$_3$H$_8$)$_2$ dimers, which we validated in subsequent molecular dynamics (MD) simulations of lower alkane systems to establish $PVT$ relationships.  This collective fitting to the FP dimer interactions was made in conjunction with the prediction of the experimental $PVT$ behavior of a dense high-pressure methane system (density of 0.5534 g/cc).  The FP data for the dimer interactions were taken from literature values, some of which are quite dated.  However, since these dimer interactions were merely used as guidance to tune in the actual pressure values of the methane system at a given density and temperature, the comparisons of MEAM predictions with the FP data (see Section \ref{sec:sec4}) should only be construed as qualitative.  The $PVT$ validations with the experimental data, together with the MEAM predictions for the bond distances, bond angles, dihedral (torsion) angles, and atomization energies of a series of alkanes and free radicals as well as the energetics of C--H and C--C bond breaking and heat of reaction for a few select reactions, are given in Section \ref{sec:sec4}.  During the parameterization process, we found the value of the  $\alpha_H^0$ parameter in the work of Baskes \cite{Bas1992} to be incorrect due to an error in the implementation of the equation for the diatomic force constant of the H--H bond.  The value of this parameter was corrected in the present work, and the corrected value appears in Table \ref{table1}.  The final sets of MEAM parameters for carbon, hydrogen, and CH is given in Tables \ref{table1} and \ref{table2}.  

\begin{sidewaystable}
 \caption{Single element MEAM parameters for carbon and hydrogen with diamond and diatomic H$_2$ reference structures, respectively. $E_\tau^0$ (\si{eV}) is the cohesive energy per atom, $R_\tau^0$ (\si{\angstrom}) is the nearest neighbor distance in the equilibrium reference structure, $\alpha_\tau^0$ is the exponential decay factor for the universal energy of Rose et~al.~\cite{Ros1984}, $A_\tau$ is the electron density scaling factor for the embedding function, $\rho_\tau^0$ is the electron density scaling factor, $\delta_\tau^a$ and $\delta_\tau^r$ are the attraction ($a^\ast>0$) and repulsion ($a^\ast\leq0$) cubic terms for the universal equation of state, $\beta_\tau^{(0-3)}$ are the exponential decay factors for the atomic electron densities, $t_\tau^{(1-3)}$ are the weighting parameters for the atomic electron densities, and $C_{min}$ and $C_{max}$ are the screening parameters for three like atoms of the element $\tau$.}
\centering 
      \begin{tabular}{c|cccccccccccccccc}
	\hline \hline \\ [-2ex]
	  Element &
	  $E_\tau^0$ & $R_\tau^0$ & $\alpha_\tau^0$ & $A_\tau$ &
	  $\rho_\tau^0$ & $\delta_\tau^a$ & $\delta_\tau^r$ &
  	  $\beta_\tau^{(0)}$ & $\beta_\tau^{(1)}$ & $\beta_\tau^{(2)}$ & $\beta_\tau^{(3)}$ &
  	  $t_\tau^{(1)}$ & $t_\tau^{(2)}$ & $t_\tau^{(3)}$ &
	  $C_{min}$ & $C_{max}$
	   \\ [0.5ex]
	\hline \\ [-2ex]
	  C &
	  \num{7.370} & \num{1.44} & \num{3.6000} & \num{0.64} &
	  \num{1.00} & \num{0.00} & \num{0.00} &
  	  \num{4.20} & \num{4.500} & \num{4.30} & \num{4.18} &
  	  \num{0.50} & \num{0.45} & \num{-3.80} &
	  \num{2.00} & \num{2.80} \\ 
	  H &
	  \num{2.363} & \num{0.74} & \num{2.0388} & \num{2.50} &
	  \num{1.80} & \num{0.00} & \num{0.05} &
  	  \num{2.72} & \num{2.045} & \num{2.25} & --  &
  	  \num{0.20} & \num{-0.40} & \num{0.00} &
	  \num{0.75} & \num{2.80} \\ [0.5ex]
	\hline
      \end{tabular}
\label{table1}
\end{sidewaystable}

\begin{table}[!htbp]
  \caption{\label{tab:alloy_pot} MEAM interaction and screening parameters for the diatomic C--H. $R_{CH}^0$ (\AA) is the first nearest neighbor distance, $\alpha_{CH}^0$  is the exponential decay factor for the universal equation of state (UEOS) of Rose  et~al.~\cite{Ros1984}, $E_{CH}^0$ (eV) is the cohesive energy per atom, $\delta_{CH}^a$  and $\delta_{CH}^r$  are the attraction ($a^\ast>0$) and repulsion ($a^\ast\leq0$) cubic terms for the UEOS, respectively, and $C_{min}$ and $C_{max}$ are the parameters for the screening factor.  The middle atom screens the other two atoms (see Section \ref{sec:sec2}).}
\centering 
    \begin{tabular}{cc}
	\hline \hline \\ [-2ex]
      Parameter & Value \\ [0.5ex]
	\hline \\ [-2ex]
      $R_{CH}^0$	& 1.02 	\\
      $\alpha_{CH}^0$	& 3.20	\\
      $E_{CH}^0$	& 2.747 	\\
      $\delta_{CH}^a$	& 0.05 	\\
      $\delta_{CH}^r$	& 0.05 	\\
      $C_{min}\left(C,C,H\right)$	&  0.445	\\
      $C_{max}\left(C,C,H\right)$	&  2.80	\\
      $C_{min}\left(C,H,C\right)$	&  2.00	\\
      $C_{max}\left(C,H,C\right)$	&  2.80	\\
      $C_{min}\left(C,H,H\right)$	&  1.50	\\
      $C_{max}\left(C,H,H\right)$	&  2.00	\\
      $C_{min}\left(H,C,H\right)$	&  0.52	\\
      $C_{max}\left(H,C,H\right)$	&  2.20	\\  [0.5ex]
      \hline
      \end{tabular}
\label{table2}
\end{table}

\section{Results \label{sec:sec4}}

\subsection{Single Molecules}

\subsubsection{Saturated Molecules }

The MEAM atomization energies of select alkanes, free radicals, and unsaturated molecules (Section \ref{unsaturated}) are given in Table \ref{table3}.  In the same table, the experimental atomization energies at 0 K \cite{Nis2011,Kar2009} are also given, along with the values calculated using the REBO and ReaxFF potentials.  The starting molecular structures were created in the Avogadro open-source molecular builder and visualization tool \cite{Han2012} and initially optimized using Avogadro's built-in Molecular Merck Force Field (MMFF94). \cite{Hal1996}  The energies of the resulting structures were then minimized using MEAM, REBO, and ReaxFF potentials and the Polak-Ribiere conjugate gradient method. \cite{Bha2000}  We utilized the parameters for carbon and hydrogen in the second-generation REBO and ReaxFF from the work of Brenner et~al.~\cite{Bre2002} and Mattsson et~al.~\cite{Mat2010}, respectively.  The REBO and ReaxFF calculations were performed on the open-source large-scale atomic/molecular massively parallel simulator (LAMMPS) software package \cite{Pli1995} developed at Sandia National Laboratories (version April 20, 2012).  The ReaxFF implementation in LAMMPS is based on the formalism introduced by Chenoweth, et~al.~\cite{Chenoweth2008} in 2008.  This implementation has been validated against the original ReaxFF serial codes. \cite{Mat2010}    

All MEAM calculations and simulations were performed on DYNAMO software (V8.7), developed by Foiles, Daw, and Baskes \cite{Foi1994} at Sandia National Laboratories.  The MEAM potential was fit to the experimental data corrected for ZPE.  Hence, the corrected value in Table III should be compared with the experimental data.  Similarly, in the second generation REBO, a ZPE correction needs to be applied to the LAMMPS-calculated energies of hydrocarbon structures. \cite{Bre2002}  The reported atomization energies in Table III reflect these corrections based on the ZPE data reported by Brenner et~al.~\cite{Bre2002} and B3LYP/6-31G** density functional theory (DFT) ZPE calculations performed as part of this work.  Furthermore, since ReaxFF was fit to the heats of formation of hydrocarbons, \cite{Van2001} it is necessary to correct the LAMMPS-calculated energies by the empirical ``heat increments'' discussed in the work of van Duin et~al.~\cite{Van2001}  We corrected the LAMMPS-calculated energies for the structures using the ReaxFF potential by calculating the differences between the ReaxFF empirical heat increments for carbon (9.489 eV) and hydrogen (2.355 eV) \cite{Van2001} and the experimental energies for carbon (7.3768 eV) and hydrogen (2.375 eV), \cite{Bre1990} and then subtracting the total difference for carbon and hydrogen atoms in the molecules from the LAMMPS-calculated energies.  The root-mean-square (RMS) error associated with the MEAM-reproduced atomization energies of the alkanes in Table \ref{table3} (0.19 eV) compares well with that of REBO (0.11 eV) and is far better than that of ReaxFF (0.99 eV).  The systems with the largest errors for MEAM contain double and triple bonds (see Section \ref{unsaturated}).  This is not surprising since MEAM is not parameterized for these systems.  

\begin{figure}[ht!]
        \centering
        \begin{subfigure}[b]{0.6\textwidth}
                \centering
                \includegraphics[width=\textwidth]{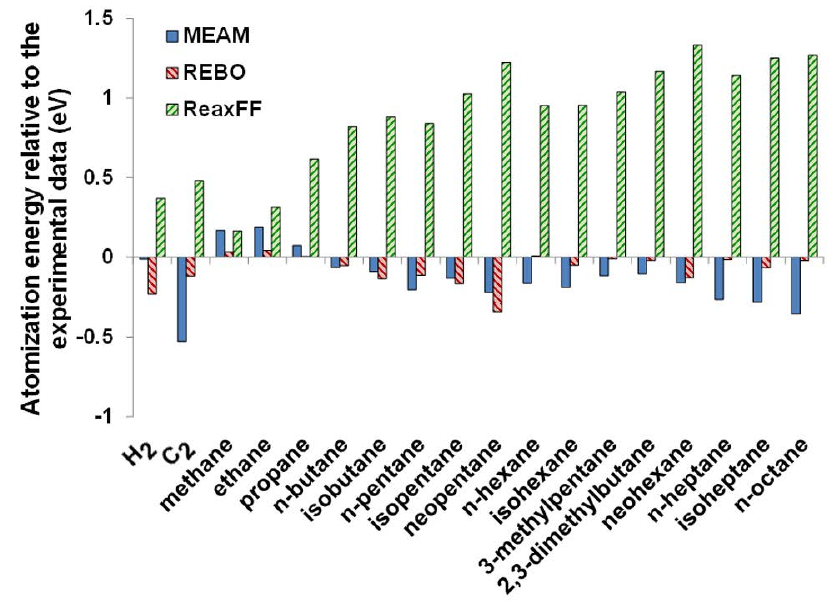}
                \caption{}
                \label{fig:fig1a}
        \end{subfigure}%
\quad
        \begin{subfigure}[b]{0.6\textwidth}
                \centering
                \includegraphics[width=\textwidth]{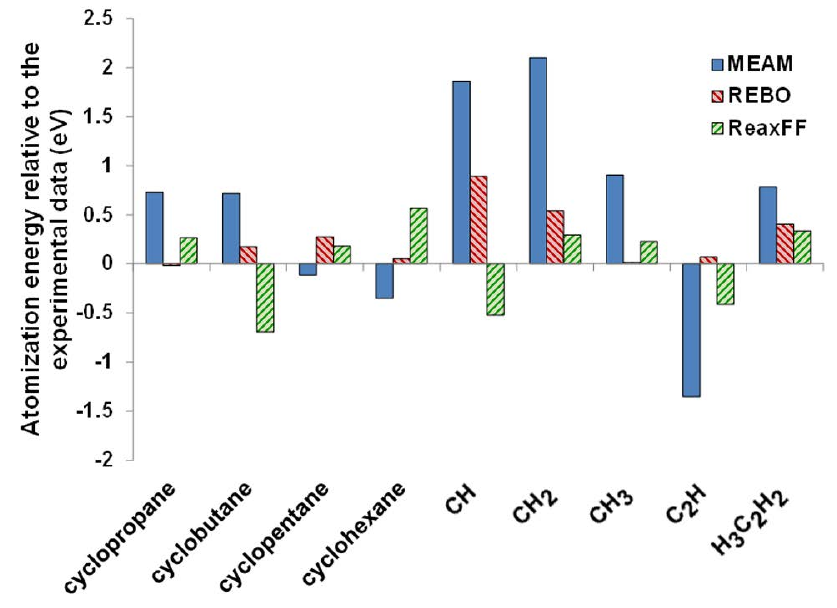}
                \caption{}
                \label{fig:fig1b}
        \end{subfigure}
        \caption{Errors associated with the corrected MEAM, REBO, and ReaxFF atomization energies of a) hydrogen and linear alkanes and b) cycloalkanes and free radicals relative to the experimental data.  The actual data are given in Table \ref{table3}.}
\label{fig:fig1}
\end{figure}

However, MEAM does give a reasonable prediction of the atomization energies of methyl (CH$_3$), ethyl (H$_3$C$_2$H$_2$), and isopropyl (CH$_3$CHCH$_3$) radicals, which are representative molecular fragments for the bond-breaking reactions that can occur in saturated hydrocarbons.  A comparison between the atomization energies relative to the experimental data for the three potentials is given in Fig.~\ref{fig:fig1}.  Of course, REBO and ReaxFF have been parameterized to a larger database of both saturated and unsaturated hydrocarbons, and therefore, they can be applied to a much larger range of systems.  Hence, the comparisons in Table \ref{table3} serve only as a guide.

\begin{figure}[ht!]
        \centering
        \begin{subfigure}[b]{0.6\textwidth}
                \centering
                \includegraphics[width=\textwidth]{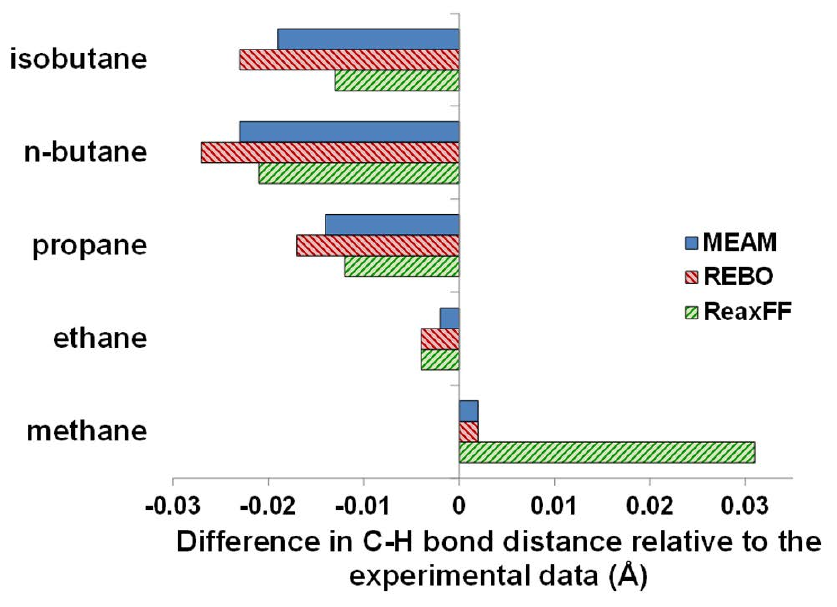}
                \caption{}
                \label{fig:fig2a}
        \end{subfigure}%
\quad
        \begin{subfigure}[b]{0.6\textwidth}
                \centering
                \includegraphics[width=\textwidth]{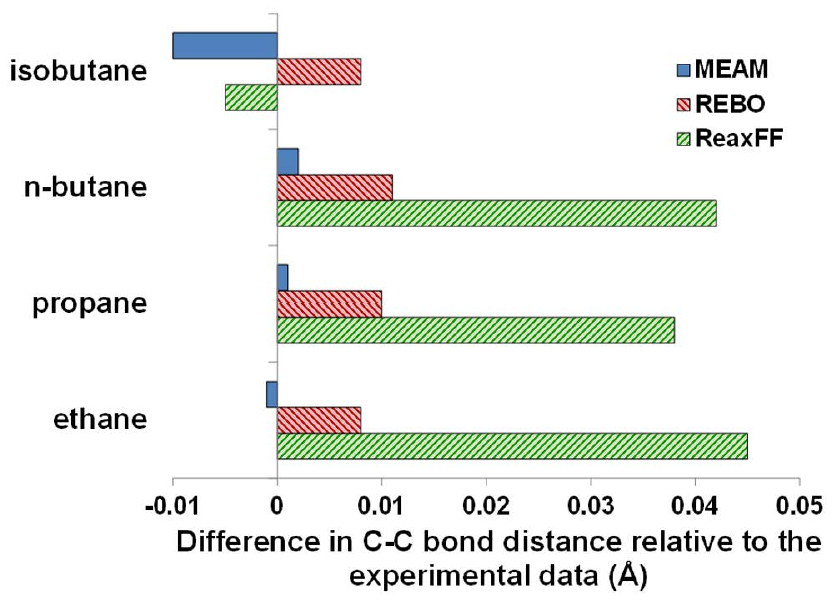}
                \caption{}
                \label{fig:fig2b}
        \end{subfigure}
        \caption{Errors associated with the MEAM-, REBO-, and ReaxFF-reproduced bond distances for the a) C--H and b) C--C bonds of select alkanes relative to the experimental data.  The actual data are given in Table \ref{table4}.}
\label{fig:fig2}
\end{figure}

\begin{figure}[ht!]
        \centering
        \begin{subfigure}[b]{0.45\textwidth}
                \centering
                \includegraphics[width=\textwidth]{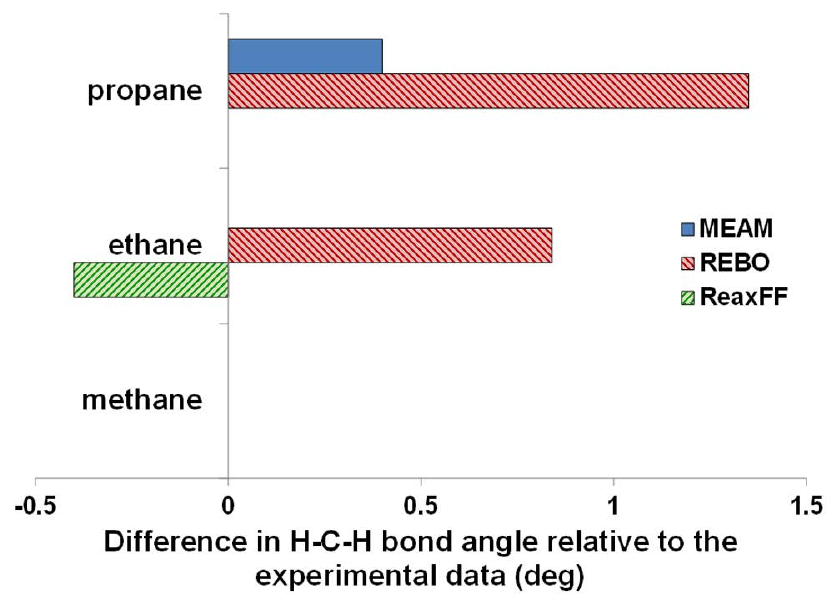}
                \caption{}
                \label{fig:fig3a}
        \end{subfigure}%
\quad
        \begin{subfigure}[b]{0.45\textwidth}
                \centering
                \includegraphics[width=\textwidth]{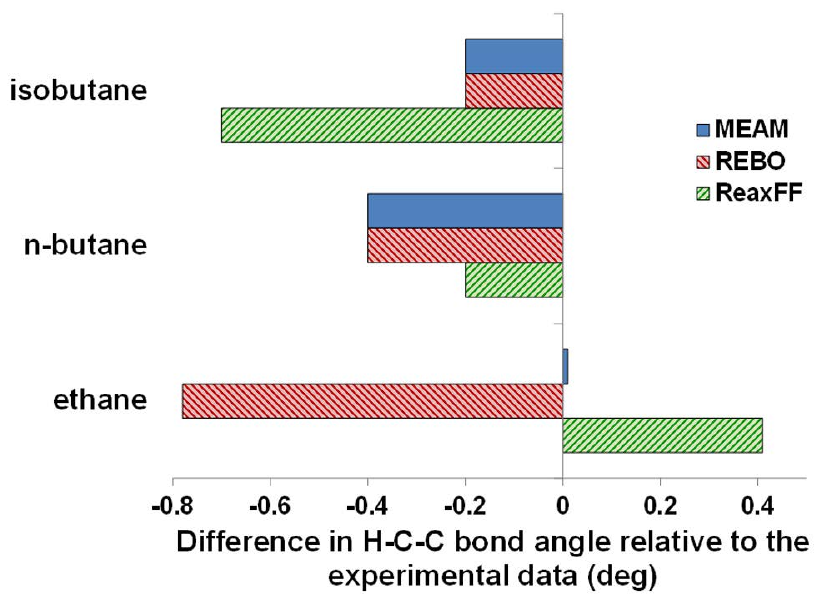}
                \caption{}
                \label{fig:fig3b}
        \end{subfigure}
\quad
        \begin{subfigure}[b]{0.45\textwidth}
                \centering
                \includegraphics[width=\textwidth]{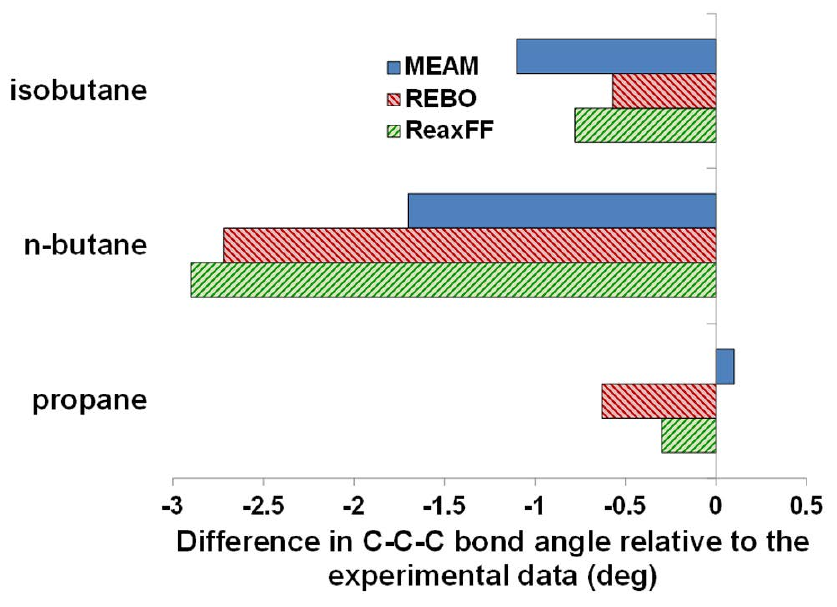}
                \caption{}
                \label{fig:fig3c}
        \end{subfigure}
        \caption{Errors associated with the MEAM-, REBO-, and ReaxFF-reproduced bond angles for the a) $\angle$H--C--H, b) $\angle$H--C--C, and c) $\angle$C--C--C angles relative to the experimental data.  The actual data are given in Table \ref{table5}.}
\label{fig:fig3}
\end{figure}

The average equilibrium bond distances and bond angles for the first three molecules in the alkane series and both isomers of butane are given in Tables \ref{table4} and \ref{table5}, respectively, where the MEAM, REBO, and ReaxFF values are compared to the experimental data. \cite{Lid2009}  The MEAM results give lower RMS errors for both bond distances and bond angles than REBO or ReaxFF.  In Figs.~\ref{fig:fig2} and \ref{fig:fig3}, the bond distances and bond angles relative to the experimental data are depicted for the three potentials.

The MEAM-predicted dihedral (torsion) angle for the gauche conformer of isobutane was predicted to be 78$^\circ$, while REBO and ReaxFF gave a prediction of 68$^\circ$ and 67$^\circ$, respectively. The experimental value of the dihedral angle for the gauche conformer of isobutane is 65$^\circ$. \cite{Lid2009}  MEAM reproduced the angle within 20\% of the experimental value, while REBO and ReaxFF reproduced it within 5\% and 3\%, respectively.

\renewcommand{\thefootnote}{\alph{footnote}}
\begin{table}[htbp!]
  \centering
\footnotesize
  \caption{Atomization energies of a homologous series of alkanes from methane to $n$-octane and their select isomers, cycloalkanes, free radicals, hydrogen, and carbon diatomic reproduced by MEAM, REBO, and ReaxFF potentials versus experimental data.  The energies of unsaturated ethylene, acetylene, and benzene are also given for illustration purposes, since the current MEAM potential has not explicitly been parameterized for such systems.  The DYNAMO- and LAMMPS-generated energies for MEAM, REBO, and ReaxFF potentials were corrected before comparison with the experimental atomization energies.}
\begin{threeparttable}[b]    
    \begin{tabular}{rrrrrrrrr}
\hline
\hline \\ [-2ex]
    \multicolumn{1}{c}{\multirow{3}[4]{*}{Molecule}} & \multicolumn{8}{c}{Atomization energy at 0 K (eV)} \\ 
\cline{2-9} \\[-2ex]
    \multicolumn{1}{c}{} & \multicolumn{1}{C{1.5cm}}{\multirow{2}[4]{*}{Expt.\footnotemark[2]}} & \multicolumn{1}{C{1.5cm}}{\multirow{2}[4]{*}{ZPE\footnotemark[4]}} & \multicolumn{2}{c}{MEAM \cite{Bas1992}} & \multicolumn{2}{c}{REBO \cite{Bre2002}} & \multicolumn{2}{c}{ReaxFF \cite{Van2001}} \\
\cline{4-9} \\ [-2ex]
    \multicolumn{1}{c}{} & \multicolumn{1}{c}{} & \multicolumn{1}{c}{} & \multicolumn{1}{C{1.5cm}}{DYNAMO} & \multicolumn{1}{C{1.5cm}}{Corr.\footnotemark[6]} & \multicolumn{1}{C{1.5cm}}{LAMMPS\footnotemark[7]} & \multicolumn{1}{C{1.5cm}}{Corr.\footnotemark[8]} & \multicolumn{1}{C{1.5cm}}{LAMMPS\footnotemark[9]} & \multicolumn{1}{C{1.5cm}}{Corr.\footnotemark[10]} \\
\hline  \\ [-2ex]
    \multicolumn{1}{c}{H$_2$} & \multicolumn{1}{c}{\textbf{4.478}} & \multicolumn{1}{c}{0.263\footnotemark[5]} & \multicolumn{1}{c}{4.726} & \multicolumn{1}{c}{\textbf{4.463}} & \multicolumn{1}{c}{4.506} & \multicolumn{1}{c}{\textbf{4.243}} & \multicolumn{1}{c}{4.804} & \multicolumn{1}{c}{\textbf{4.845}} \\ 
    \multicolumn{1}{c}{C$_2$} & \multicolumn{1}{c}{\textbf{6.219}} & \multicolumn{1}{c}{0.111\footnotemark[5]} & \multicolumn{1}{c}{5.804} & \multicolumn{1}{c}{\textbf{5.693}} & \multicolumn{1}{c}{6.210} & \multicolumn{1}{c}{\textbf{6.099}} & \multicolumn{1}{c}{10.902} & \multicolumn{1}{c}{\textbf{6.697}} \\
    \multicolumn{1}{l}{} & \multicolumn{1}{c}{\textbf{}} & \multicolumn{1}{c}{} & \multicolumn{1}{c}{} & \multicolumn{1}{c}{\textbf{}} & \multicolumn{1}{c}{} & \multicolumn{1}{c}{\textbf{}} & \multicolumn{1}{c}{} & \multicolumn{1}{c}{\textbf{}} \\  [-1ex]
    \multicolumn{1}{l}{\underline{Alkanes}} & \multicolumn{1}{c}{\textbf{}} & \multicolumn{1}{c}{} & \multicolumn{1}{c}{} & \multicolumn{1}{c}{\textbf{}} & \multicolumn{1}{c}{} & \multicolumn{1}{c}{\textbf{}} & \multicolumn{1}{c}{} & \multicolumn{1}{c}{\textbf{}} \\
    \multicolumn{1}{c}{methane} & \multicolumn{1}{c}{\textbf{17.018}} & \multicolumn{1}{c}{1.135} & \multicolumn{1}{c}{18.319} & \multicolumn{1}{c}{\textbf{17.184}} & \multicolumn{1}{c}{18.185} & \multicolumn{1}{c}{\textbf{17.050}} & \multicolumn{1}{c}{19.202} & \multicolumn{1}{c}{\textbf{17.181}} \\
    \multicolumn{1}{c}{ethane} & \multicolumn{1}{c}{\textbf{28.885}} & \multicolumn{1}{c}{1.921} & \multicolumn{1}{c}{30.991} & \multicolumn{1}{c}{\textbf{29.070}} & \multicolumn{1}{c}{30.846} & \multicolumn{1}{c}{\textbf{28.925}} & \multicolumn{1}{c}{33.279} & \multicolumn{1}{c}{\textbf{29.196}} \\
    \multicolumn{1}{c}{propane} & \multicolumn{1}{c}{\textbf{40.880}} & \multicolumn{1}{c}{2.706} & \multicolumn{1}{c}{43.658} & \multicolumn{1}{c}{\textbf{40.952}} & \multicolumn{1}{c}{43.589} & \multicolumn{1}{c}{\textbf{40.883}} & \multicolumn{1}{c}{47.640} & \multicolumn{1}{c}{\textbf{41.495}} \\
    \multicolumn{1}{c}{$n$-butane} & \multicolumn{1}{c}{\textbf{52.896}} & \multicolumn{1}{c}{3.492} & \multicolumn{1}{c}{56.322} & \multicolumn{1}{c}{\textbf{52.830}} & \multicolumn{1}{c}{56.332} & \multicolumn{1}{c}{\textbf{52.840}} & \multicolumn{1}{c}{61.921} & \multicolumn{1}{c}{\textbf{53.714}} \\
    \multicolumn{1}{c}{isobutane} & \multicolumn{1}{c}{\textbf{52.977}} & \multicolumn{1}{c}{3.492} & \multicolumn{1}{c}{56.377} & \multicolumn{1}{c}{\textbf{52.885}} & \multicolumn{1}{c}{56.331} & \multicolumn{1}{c}{\textbf{52.839}} & \multicolumn{1}{c}{62.063} & \multicolumn{1}{c}{\textbf{53.856}} \\
    \multicolumn{1}{c}{$n$-pentane} & \multicolumn{1}{c}{\textbf{64.915}} & \multicolumn{1}{c}{4.278} & \multicolumn{1}{c}{68.985} & \multicolumn{1}{c}{\textbf{64.707}} & \multicolumn{1}{c}{69.076} & \multicolumn{1}{c}{\textbf{64.798}} & \multicolumn{1}{c}{76.022} & \multicolumn{1}{c}{\textbf{65.753}} \\
    \multicolumn{1}{c}{isopentane} & \multicolumn{1}{c}{\textbf{64.964}} & \multicolumn{1}{c}{4.278} & \multicolumn{1}{c}{69.107} & \multicolumn{1}{c}{\textbf{64.829}} & \multicolumn{1}{c}{69.073} & \multicolumn{1}{c}{\textbf{64.795}} & \multicolumn{1}{c}{76.260} & \multicolumn{1}{c}{\textbf{65.991}} \\
    \multicolumn{1}{c}{neopentane} & \multicolumn{1}{c}{\textbf{65.123}} & \multicolumn{1}{c}{4.278} & \multicolumn{1}{c}{69.177} & \multicolumn{1}{c}{\textbf{64.899}} & \multicolumn{1}{c}{69.061} & \multicolumn{1}{c}{\textbf{64.783}} & \multicolumn{1}{c}{76.614} & \multicolumn{1}{c}{\textbf{66.345}} \\
    \multicolumn{1}{c}{$n$-hexane} & \multicolumn{1}{c}{\textbf{76.922}} & \multicolumn{1}{c}{4.892\footnotemark[5]} & \multicolumn{1}{c}{81.648} & \multicolumn{1}{c}{\textbf{76.756}} & \multicolumn{1}{c}{81.819} & \multicolumn{1}{c}{\textbf{76.929}} & \multicolumn{1}{c}{90.204} & \multicolumn{1}{c}{\textbf{77.873}} \\
    \multicolumn{1}{c}{isohexane} & \multicolumn{1}{c}{\textbf{76.975}} & \multicolumn{1}{c}{4.896\footnotemark[5]} & \multicolumn{1}{c}{81.680} & \multicolumn{1}{c}{\textbf{76.784}} & \multicolumn{1}{c}{81.817} & \multicolumn{1}{c}{\textbf{76.921}} & \multicolumn{1}{c}{90.259} & \multicolumn{1}{c}{\textbf{77.928}} \\
    \multicolumn{1}{c}{3-methylpentane} & \multicolumn{1}{c}{\textbf{76.946}} & \multicolumn{1}{c}{4.885\footnotemark[5]} & \multicolumn{1}{c}{81.712} & \multicolumn{1}{c}{\textbf{76.827}} & \multicolumn{1}{c}{81.817} & \multicolumn{1}{c}{\textbf{76.932}} & \multicolumn{1}{c}{90.312} & \multicolumn{1}{c}{\textbf{77.981}} \\
    \multicolumn{1}{c}{2,3-dimethylbutane} & \multicolumn{1}{c}{\textbf{76.970}} & \multicolumn{1}{c}{4.867\footnotemark[5]} & \multicolumn{1}{c}{81.730} & \multicolumn{1}{c}{\textbf{76.863}} & \multicolumn{1}{c}{81.815} & \multicolumn{1}{c}{\textbf{76.948}} & \multicolumn{1}{c}{90.467} & \multicolumn{1}{c}{\textbf{78.136}} \\
    \multicolumn{1}{c}{neohexane} & \multicolumn{1}{c}{\textbf{77.060}} & \multicolumn{1}{c}{4.876\footnotemark[5]} & \multicolumn{1}{c}{81.772} & \multicolumn{1}{c}{\textbf{76.896}} & \multicolumn{1}{c}{81.804} & \multicolumn{1}{c}{\textbf{76.928}} & \multicolumn{1}{c}{90.721} & \multicolumn{1}{c}{\textbf{78.390}} \\
    \multicolumn{1}{c}{$n$-heptane} & \multicolumn{1}{c}{\textbf{88.957}\footnotemark[3]} & \multicolumn{1}{c}{5.623\footnotemark[5]} & \multicolumn{1}{c}{94.311} & \multicolumn{1}{c}{\textbf{88.688}} & \multicolumn{1}{c}{94.562} & \multicolumn{1}{c}{\textbf{88.939}} & \multicolumn{1}{c}{104.489} & \multicolumn{1}{c}{\textbf{90.096}} \\
    \multicolumn{1}{c}{isoheptane} & \multicolumn{1}{c}{\textbf{89.008}\footnotemark[3]} & \multicolumn{1}{c}{5.622\footnotemark[5]} & \multicolumn{1}{c}{94.346} & \multicolumn{1}{c}{\textbf{88.724}} & \multicolumn{1}{c}{94.560} & \multicolumn{1}{c}{\textbf{88.938}} & \multicolumn{1}{c}{104.652} & \multicolumn{1}{c}{\textbf{90.259}} \\
    \multicolumn{1}{c}{$n$-octane} & \multicolumn{1}{c}{\textbf{100.971}\footnotemark[3]} & \multicolumn{1}{c}{6.359\footnotemark[5]} & \multicolumn{1}{c}{106.975} & \multicolumn{1}{c}{\textbf{100.616}} & \multicolumn{1}{c}{107.306} & \multicolumn{1}{c}{\textbf{100.947}} & \multicolumn{1}{c}{118.692} & \multicolumn{1}{c}{\textbf{102.237}} \\
    \multicolumn{1}{c}{RMS Error\footnotemark[1]:} & \multicolumn{1}{c}{-} & \multicolumn{1}{c}{-} & \multicolumn{1}{c}{-} & \multicolumn{1}{c}{0.19} & \multicolumn{1}{c}{-} & \multicolumn{1}{c}{0.11} & \multicolumn{1}{c}{-} & \multicolumn{1}{c}{0.99} \\
    \multicolumn{1}{l}{} & \multicolumn{1}{c}{\textbf{}} & \multicolumn{1}{c}{} & \multicolumn{1}{c}{} & \multicolumn{1}{c}{\textbf{}} & \multicolumn{1}{c}{} & \multicolumn{1}{c}{\textbf{}} & \multicolumn{1}{c}{} & \multicolumn{1}{c}{\textbf{}} \\  [-1ex]
    \multicolumn{1}{l}{\underline{Alkenes}} & \multicolumn{1}{c}{\textbf{}} & \multicolumn{1}{c}{} & \multicolumn{1}{c}{} & \multicolumn{1}{c}{\textbf{}} & \multicolumn{1}{c}{} & \multicolumn{1}{c}{\textbf{}} & \multicolumn{1}{c}{} & \multicolumn{1}{c}{\textbf{}} \\
    \multicolumn{1}{c}{ethene (ethylene)} & \multicolumn{1}{c}{\textbf{23.066}} & \multicolumn{1}{c}{1.303} & \multicolumn{1}{c}{22.955} & \multicolumn{1}{c}{\textbf{21.652}} & \multicolumn{1}{c}{24.528} & \multicolumn{1}{c}{\textbf{23.225}} & \multicolumn{1}{c}{27.183} & \multicolumn{1}{c}{\textbf{23.059}} \\
    \multicolumn{1}{c}{} & \multicolumn{1}{c}{\textbf{}} & \multicolumn{1}{c}{} & \multicolumn{1}{c}{} & \multicolumn{1}{c}{\textbf{}} & \multicolumn{1}{c}{} & \multicolumn{1}{c}{\textbf{}} & \multicolumn{1}{c}{} & \multicolumn{1}{c}{\textbf{}} \\ [-1ex]
    \multicolumn{1}{l}{\underline{Alkynes}} & \multicolumn{1}{c}{\textbf{}} & \multicolumn{1}{c}{} & \multicolumn{1}{c}{} & \multicolumn{1}{c}{\textbf{}} & \multicolumn{1}{c}{} & \multicolumn{1}{c}{\textbf{}} & \multicolumn{1}{c}{} & \multicolumn{1}{c}{\textbf{}} \\
    \multicolumn{1}{c}{ethyne (acetylene)} & \multicolumn{1}{c}{\textbf{16.857}} & \multicolumn{1}{c}{0.686} & \multicolumn{1}{c}{14.552} & \multicolumn{1}{c}{\textbf{13.866}} & \multicolumn{1}{c}{17.565} & \multicolumn{1}{c}{\textbf{16.879}} & \multicolumn{1}{c}{19.952} & \multicolumn{1}{c}{\textbf{15.7874}} \\
    \multicolumn{1}{l}{} & \multicolumn{1}{c}{\textbf{}} & \multicolumn{1}{c}{} & \multicolumn{1}{c}{} & \multicolumn{1}{c}{\textbf{}} & \multicolumn{1}{c}{} & \multicolumn{1}{c}{\textbf{}} & \multicolumn{1}{c}{} & \multicolumn{1}{c}{\textbf{}} \\  [-1ex]
    \multicolumn{1}{l}{\underline{Cycloalkanes}} & \multicolumn{1}{c}{\textbf{}} & \multicolumn{1}{c}{} & \multicolumn{1}{c}{} & \multicolumn{1}{c}{\textbf{}} & \multicolumn{1}{c}{} & \multicolumn{1}{c}{\textbf{}} & \multicolumn{1}{c}{} & \multicolumn{1}{c}{\textbf{}} \\ 
    \multicolumn{1}{c}{cyclopropane} & \multicolumn{1}{c}{\textbf{34.818}} & \multicolumn{1}{c}{2.089} & \multicolumn{1}{c}{37.636} & \multicolumn{1}{c}{\textbf{35.547}} & \multicolumn{1}{c}{36.889} & \multicolumn{1}{c}{\textbf{34.800}} & \multicolumn{1}{c}{41.270} & \multicolumn{1}{c}{\textbf{35.084}} \\
    \multicolumn{1}{c}{cyclobutane} & \multicolumn{1}{c}{\textbf{46.848}} & \multicolumn{1}{c}{2.875} & \multicolumn{1}{c}{50.441} & \multicolumn{1}{c}{\textbf{47.566}} & \multicolumn{1}{c}{49.898} & \multicolumn{1}{c}{\textbf{47.023}} & \multicolumn{1}{c}{54.402} & \multicolumn{1}{c}{\textbf{46.154}} \\
    \multicolumn{1}{c}{cyclopentane} & \multicolumn{1}{c}{\textbf{59.707}} & \multicolumn{1}{c}{3.660} & \multicolumn{1}{c}{63.252} & \multicolumn{1}{c}{\textbf{59.592}} & \multicolumn{1}{c}{63.643} & \multicolumn{1}{c}{\textbf{59.983}} & \multicolumn{1}{c}{70.197} & \multicolumn{1}{c}{\textbf{59.887}} \\
    \multicolumn{1}{c}{cyclohexane} & \multicolumn{1}{c}{\textbf{71.963}} & \multicolumn{1}{c}{4.446} & \multicolumn{1}{c}{76.059} & \multicolumn{1}{c}{\textbf{71.613}} & \multicolumn{1}{c}{76.460} & \multicolumn{1}{c}{\textbf{72.014}} & \multicolumn{1}{c}{84.901} & \multicolumn{1}{c}{\textbf{72.529}} \\
    \multicolumn{1}{c}{RMS Error\footnotemark[1]:} & \multicolumn{1}{c}{\textbf{-}} & \multicolumn{1}{c}{-} & \multicolumn{1}{c}{-} & \multicolumn{1}{c}{0.54} & \multicolumn{1}{c}{-} & \multicolumn{1}{c}{0.17} & \multicolumn{1}{c}{-} & \multicolumn{1}{c}{0.48} \\
    \multicolumn{1}{c}{} & \multicolumn{1}{c}{\textbf{}} & \multicolumn{1}{c}{} & \multicolumn{1}{c}{} & \multicolumn{1}{c}{} & \multicolumn{1}{c}{} & \multicolumn{1}{c}{} & \multicolumn{1}{c}{} & \multicolumn{1}{c}{} \\  [-1ex]
    \multicolumn{1}{l}{\underline{Aromatics}} & \multicolumn{1}{c}{\textbf{}} & \multicolumn{1}{c}{} & \multicolumn{1}{c}{} & \multicolumn{1}{c}{\textbf{}} & \multicolumn{1}{c}{} & \multicolumn{1}{c}{\textbf{}} & \multicolumn{1}{c}{} & \multicolumn{1}{c}{\textbf{}} \\
    \multicolumn{1}{c}{benzene} & \multicolumn{1}{c}{\textbf{56.62}} & \multicolumn{1}{c}{2.594} & \multicolumn{1}{c}{52.308} & \multicolumn{1}{c}{\textbf{49.714}} & \multicolumn{1}{c}{60.231} & \multicolumn{1}{c}{\textbf{57.637}} & \multicolumn{1}{c}{68.236} & \multicolumn{1}{c}{\textbf{55.742}} \\
    \multicolumn{1}{l}{} & \multicolumn{1}{c}{\textbf{}} & \multicolumn{1}{c}{} & \multicolumn{1}{c}{} & \multicolumn{1}{c}{\textbf{}} & \multicolumn{1}{c}{} & \multicolumn{1}{c}{\textbf{}} & \multicolumn{1}{c}{} & \multicolumn{1}{c}{\textbf{}} \\  [-1ex]
    \multicolumn{1}{l}{\underline{Radicals}} & \multicolumn{1}{c}{\textbf{}} & \multicolumn{1}{c}{} & \multicolumn{1}{c}{} & \multicolumn{1}{c}{\textbf{}} & \multicolumn{1}{c}{} & \multicolumn{1}{c}{\textbf{}} & \multicolumn{1}{c}{} & \multicolumn{1}{c}{\textbf{}} \\
    \multicolumn{1}{c}{CH} & \multicolumn{1}{c}{\textbf{3.469}} & \multicolumn{1}{c}{0.165\footnotemark[5]} & \multicolumn{1}{c}{5.493} & \multicolumn{1}{c}{\textbf{5.328}} & \multicolumn{1}{c}{4.526} & \multicolumn{1}{c}{\textbf{4.361}} & \multicolumn{1}{c}{5.029} & \multicolumn{1}{c}{\textbf{2.947}} \\
    \multicolumn{1}{c}{CH$_2$} & \multicolumn{1}{c}{\textbf{7.410}} & \multicolumn{1}{c}{0.517} & \multicolumn{1}{c}{10.027} & \multicolumn{1}{c}{\textbf{9.510}} & \multicolumn{1}{c}{8.469} & \multicolumn{1}{c}{\textbf{7.952}} & \multicolumn{1}{c}{9.766} & \multicolumn{1}{c}{\textbf{7.704}} \\
    \multicolumn{1}{c}{CH$_3$} & \multicolumn{1}{c}{\textbf{12.534}} & \multicolumn{1}{c}{0.826} & \multicolumn{1}{c}{14.265} & \multicolumn{1}{c}{\textbf{13.439}} & \multicolumn{1}{c}{13.375} & \multicolumn{1}{c}{\textbf{12.549}} & \multicolumn{1}{c}{14.806} & \multicolumn{1}{c}{\textbf{12.764}} \\
    \multicolumn{1}{c}{C$_2$H} & \multicolumn{1}{c}{\textbf{11.125}} & \multicolumn{1}{c}{0.377} & \multicolumn{1}{c}{10.149} & \multicolumn{1}{c}{\textbf{9.772}} & \multicolumn{1}{c}{11.572} & \multicolumn{1}{c}{\textbf{11.195}} & \multicolumn{1}{c}{14.897} & \multicolumn{1}{c}{\textbf{10.712}} \\
    \multicolumn{1}{c}{H$_3$C$_2$H$_2$} & \multicolumn{1}{c}{\textbf{24.572}} & \multicolumn{1}{c}{1.612} & \multicolumn{1}{c}{26.967} & \multicolumn{1}{c}{\textbf{25.355}} & \multicolumn{1}{c}{26.588} & \multicolumn{1}{c}{\textbf{24.976}} & \multicolumn{1}{c}{29.012} & \multicolumn{1}{c}{\textbf{24.908}} \\
    \multicolumn{1}{c}{CH$_3$CHCH$_3$} & \multicolumn{1}{c}{\textbf{36.676}} & \multicolumn{1}{c}{2.405} & \multicolumn{1}{c}{39.672} & \multicolumn{1}{c}{\textbf{37.267}} & \multicolumn{1}{c}{39.554} & \multicolumn{1}{c}{\textbf{37.148}} & \multicolumn{1}{c}{43.027} & \multicolumn{1}{c}{\textbf{36.861}} \\   
    \multicolumn{1}{c}{RMS Error\footnotemark[1]:} & \multicolumn{1}{c}{\textbf{-}} & \multicolumn{1}{c}{-} & \multicolumn{1}{c}{-} & \multicolumn{1}{c}{1.38} & \multicolumn{1}{c}{-} & \multicolumn{1}{c}{0.50} & \multicolumn{1}{c}{-} & \multicolumn{1}{c}{0.35} \\ [0.5ex]
\hline
    \end{tabular}%
        \begin{tablenotes}
\scriptsize
            \item [a] Root-Mean-Square Error.
            \item [b] From the NIST Computational Chemistry Comparison and Benchmark Database. \cite{Nis2011}
 	 \item [c] From Karton et~al.~\cite{Kar2009}
 	 \item [d] Zero-point energies are taken from Brenner et~al.~\cite{Bre2002} if not indicated otherwise.
 	 \item [e] B3LYP/6-31G** ZPEs with a scaling factor of 0.95 applied to account for method limitations and anharmonicity.
 	 \item [f] Corrections were made by subtracting the zero-point energies from the DYNAMO-calculated\cite{Foi1994} energies.
 	 \item [g] C and H parameters in LAMMPS \cite{Pli1995} (version April 20, 2012) were taken from Brenner et~al.~\cite{Bre2002}
 	 \item [h] Corrections were made by subtracting the zero-point energies from the LAMMPS-calculated\cite{Pli1995} energies.
 	 \item [i] C and H parameters in LAMMPS \cite{Pli1995} (version April 20, 2012) were taken from Mattsson et~al.~\cite{Mat2010}
 	 \item [j] Corrections were made by calculating the differences in the ``heat increments'' for carbon (9.489 eV) and hydrogen (2.355 eV), as described in the work of  van Duin et~al.~\cite{Van2001} from the experimental energies for carbon (7.3768 eV) and hydrogen (2.375 eV), as reported by Brenner et~al.~\cite{Bre1990}, respectively, and then subtracting the total difference for carbon and hydrogen atoms in the molecules from the LAMMPS-calculated\cite{Pli1995} energies.
        \end{tablenotes}
\end{threeparttable}
  \label{table3}%
\end{table}%

\begin{table}[htbp]
  \centering
  \caption{Average equilibrium C--H and C--C bond distances for select alkanes after energy minimization of the molecular structures using the MEAM, REBO, and ReaxFF potentials.  The results are compared to the experimental data. \cite{Lid2009}}
\begin{threeparttable}[b]    
\begin{tabular}{rC{1.45cm}C{1.45cm}C{1.45cm}C{1.45cm}|C{1.45cm}C{1.45cm}C{1.45cm}C{1.45cm}}
    \hline
    \hline  \\[-2ex]
    \multicolumn{1}{c}{\multirow{3}[4]{*}{Molecule}} & \multicolumn{8}{c}{Bond distance (\AA)} \\

          & \multicolumn{4}{c|}{C--H}       & \multicolumn{4}{c}{C--C} \\
\cline{2-9}  \\[-2ex]
          & Expt.\tnote{a}   & MEAM  & REBO  & ReaxFF & Expt.\tnote{a}   & MEAM  & REBO  & ReaxFF \\
    \hline  \\[-2ex]
    \multicolumn{1}{c}{methane} & 1.087 & 1.089 & 1.089 & 1.118 & -     & -     & -     & - \\
    \multicolumn{1}{c}{ethane} & 1.094 & 1.092 & 1.090  & 1.090  & 1.535 & 1.534 & 1.543 & 1.534 \\
    \multicolumn{1}{c}{propane} & 1.107 & 1.093 & 1.090  & 1.110 & 1.532 & 1.533 & 1.542 & 1.542 \\
    \multicolumn{1}{c}{$n$-butane} & 1.117 & 1.094 & 1.090  & 1.110 & 1.531 & 1.533 & 1.542 & 1.556 \\
    \multicolumn{1}{c}{isobutane} & 1.113 & 1.094 & 1.090  & 1.112   & 1.535 & 1.525 & 1.543 & 1.545 \\
    \hline  \\[-2ex]
    \multicolumn{1}{c}{RMS Error\tnote{b}} & - & 0.015 & 0.018  & 0.019   & - & 0.005 & 0.009 & 0.036 \\   [0.5ex]
    \hline
    \end{tabular}%
        \begin{tablenotes}
            \item [a] From Lide. \cite{Lid2009}
            \item [b] Root-Mean-Square Error.
        \end{tablenotes}
\end{threeparttable}
  \label{table4}%
\end{table}%

\begin{sidewaystable}
  \centering
  \caption{Average equilibrium $\angle$H--C--H, $\angle$H--C--C, and $\angle$C--C--C bond angles for select alkanes after energy minimization of the molecular structures using the MEAM, REBO, and ReaxFF potentials.  The results are compared to the experimental data. \cite{Lid2009}}
\begin{threeparttable}[b]    
\resizebox{\linewidth}{!}{%
\tabcolsep=2pt
    \begin{tabular}{rC{1.45cm}C{1.45cm}C{1.45cm}C{1.5cm}|C{1.45cm}C{1.45cm}C{1.45cm}C{1.5cm}|C{1.45cm}C{1.45cm}C{1.45cm}C{1.5cm}}
    \hline
    \hline  \\[-2ex]
    \multicolumn{1}{c}{\multirow{3}[4]{*}{Molecule}} & \multicolumn{12}{c}{Bond angle ($^\circ$)} \\
          & \multicolumn{4}{c|}{$\angle$H--C--H}     & \multicolumn{4}{c|}{$\angle$H--C--C}     & \multicolumn{4}{c}{$\angle$C--C--C} \\
\cline{2-13}  \\[-2ex]
          & Expt.\tnote{a}   & MEAM  & REBO  & ReaxFF & Expt.\tnote{a}   & MEAM  & REBO  & ReaxFF & Expt.\tnote{a}   & MEAM  & REBO  & ReaxFF \\
    \hline  \\[-2ex]
    \multicolumn{1}{c}{methane} & 109.47 & 109.47 & 109.47 & 109.47 & -     & -     & -     & -     & -     & -     & -     & - \\
    \multicolumn{1}{c}{ethane} & 107.70 & 107.70 & 108.54 & 107.30 & 111.17 & 111.18 & 110.39 & 111.58 & -     & -     & -     & - \\
    \multicolumn{1}{c}{propane} & 107.00   & 107.40 & 108.35 & 107.00   & N/A   & 110.90 & 110.20 & 110.91 & 111.70 & 111.80 & 111.07 & 111.40 \\
    \multicolumn{1}{c}{$n$-butane} & N/A   & 107.50 & 108.22 & 106.89 & 111.00   & 110.60 & 110.08 & 110.80 & 113.80 & 112.10 & 111.08 & 110.90 \\
    \multicolumn{1}{c}{isobutane} & N/A   & 107.40 & 108.54 & 107.90 & 111.40 & 111.20 & 110.39 & 110.70 & 110.80 & 109.70 & 110.23 & 110.02 \\   [0.5ex]
    \hline  \\[-2ex]
    \multicolumn{1}{c}{RMS Error\tnote{b}} & -   & 0.23 & 0.92 & 0.23 & - & 0.26 & 0.91 & 0.48 & -  & 1.17 & 1.65 & 1.74 \\
    \hline
    \end{tabular}}%
        \begin{tablenotes}%
            \item [a] From Lide. \cite{Lid2009}
            \item [b] Root-Mean-Square Error.
        \end{tablenotes}
\end{threeparttable}
  \label{table5}%
\end{sidewaystable}

\subsubsection{Unsaturated Molecules \label{unsaturated}}

The MEAM potential at its current state does not predict the energetics and structures of unsaturated molecules correctly.  To make the reader aware of this fact, the atomization energies of select unsaturated molecules are presented in Table \ref{table3}.  It is our intention to caution the readers against using the current MEAM potential for calculations related to unsaturated hydrocarbons.  The reproduction of physical properties associated with these molecules is a subject of future research and requires a modification to the MEAM potential and introduction of a bond order formalism.  The REBO and ReaxFF potentials have both neighbor counting mechanisms and a decision step that determines the bond order and distinguishes between a radical case and an unsaturated molecule case.

To test the prediction of planar structure and handling of ``over-saturation'' by the three potentials (MEAM, REBO, and ReaxFF),  two atomic configurations were selected and energy-minimized, respectively: 1) CH$_3$ in non-planar configuration with an initial  $\angle$H--C--H of 109.5$^\circ$ (to validate the formation of planar structure after energy minimization and a final  $\angle$H--C--H of 120$^\circ$), and 2) CH$_4$+2H with the two extra hydrogen atoms on either side of the methane molecule and very close to it (to validate the repulsion of the two extra hydrogen atoms and formation of a hydrogen molecule away from the methane molecule).  Correct prediction of the resulting molecular species and their geometries is crucial for reliably simulating reactions involving free radicals.  MEAM and REBO both predicted a planar CH$_3$ structure with a final  $\angle$H--C--H of 120$^\circ$ after energy minimization, while ReaxFF gave a slightly non-planar structure with a final  $\angle$H--C--H of 117.4$^\circ$.  Furthermore, for the CH$_4$+2H structure, both MEAM and REBO predicted the formation of a hydrogen molecule away from an equilibrated methane molecule, while ReaxFF minimized the structure to a non-equilibrium configuration.  

\begin{figure}[ht]
\centering
\includegraphics[width = 0.5\textwidth]{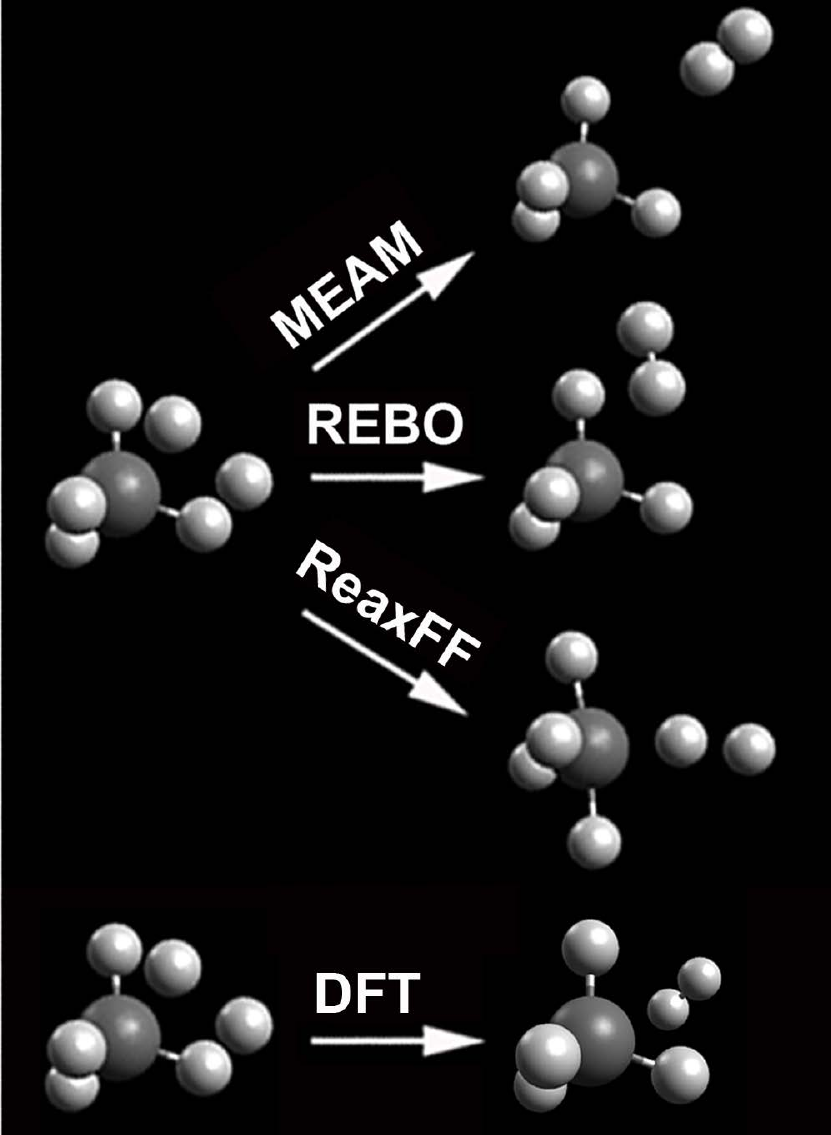}
        \caption{Results of the ``over-saturation'' test for the CH$_4$+2H atomic configuration.  Energy-minimized structures using the MEAM, REBO, and ReaxFF potentials as well as the DFT-predicted structure are presented.}
\label{fig:fig4}
\end{figure}

To compare these results with the FP data, density functional theory (DFT) calculation was run on the CH$_4$+2H structure using the aug-cc-pVDZ basis set and m062x exchange functional.  The initial and final structures for the CH$_4$+2H atomic configuration calculated by MEAM, REBO, ReaxFF, and DFT are given in Fig.~\ref{fig:fig4}.  The mechanism by which over-saturation is handled by the three potentials and DFT is related to finding the most energetically favorable atomic configuration. In the MEAM potential, as the number of neighbors to an atom increases, the background electron density increases. This causes the embedding function to increase, and, therefore, highly over-saturated molecules are not favored energetically. Similar arguments can be made for the other potentials and DFT. In the case of the CH$_4$+2H atomic configuration, the expected structure after energy minimization is one methane and one hydrogen molecule away from each other.

\clearpage

\subsection{Diatomic Molecules}

\begin{figure}[ht!]
        \centering
        \begin{subfigure}[b]{0.45\textwidth}
                \centering
                \includegraphics[width=\textwidth]{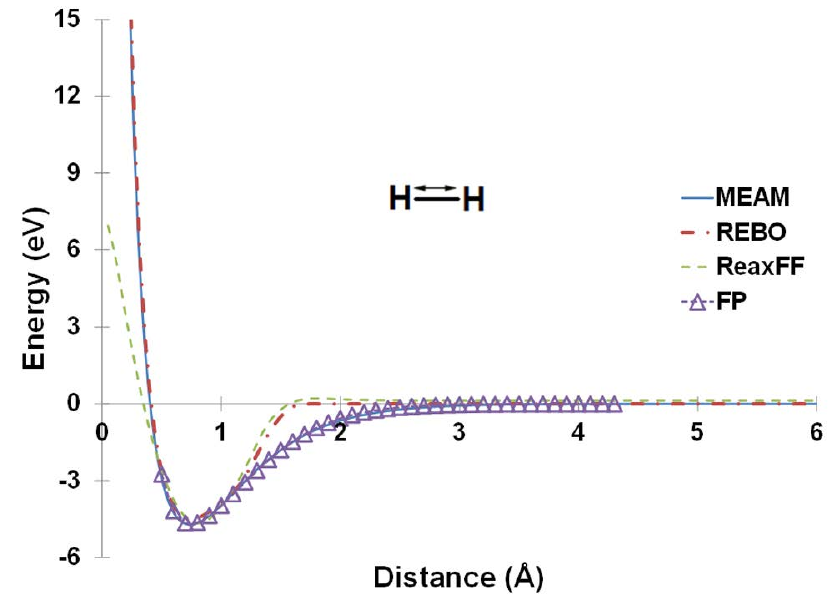}
                \caption{}
                \label{fig:fig5a}
        \end{subfigure}%
\quad
        \begin{subfigure}[b]{0.45\textwidth}
                \centering
                \includegraphics[width=\textwidth]{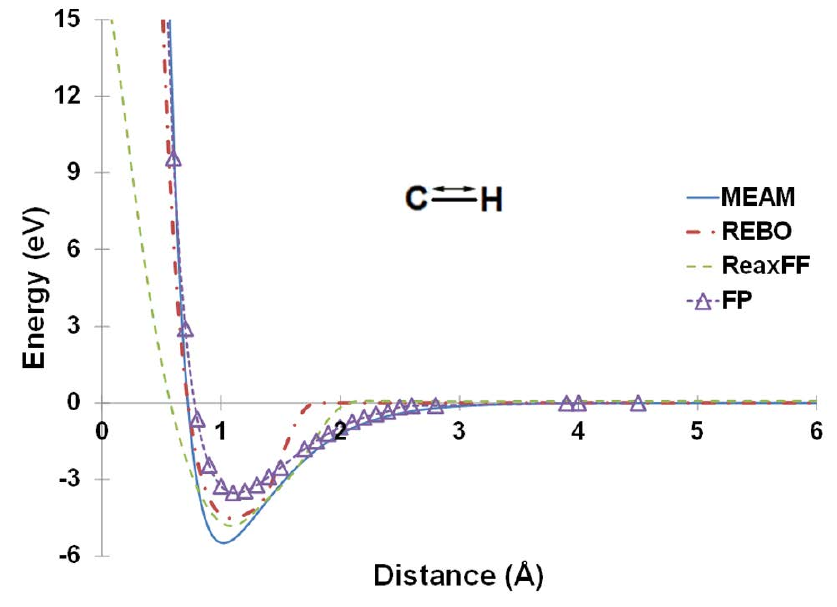}
                \caption{}
                \label{fig:fig5b}
        \end{subfigure}
\quad
        \begin{subfigure}[b]{0.45\textwidth}
                \centering
                \includegraphics[width=\textwidth]{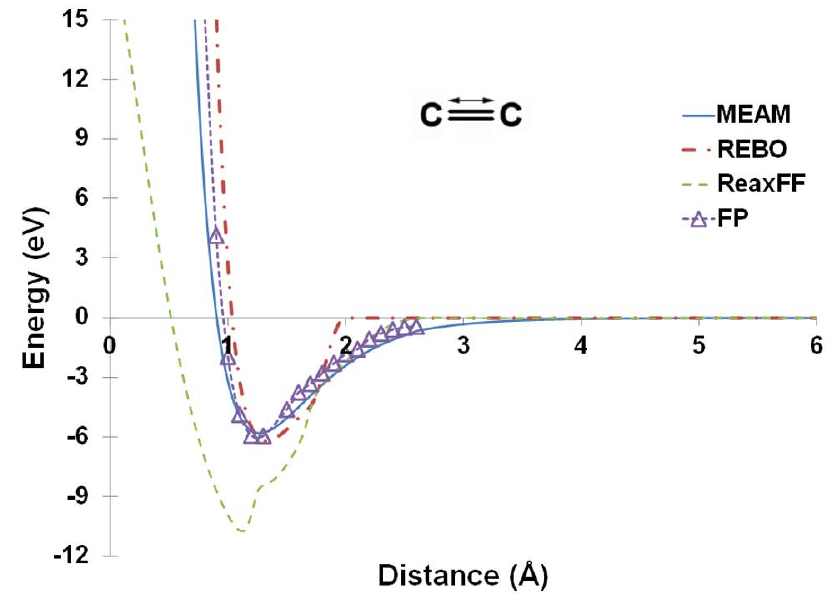}
                \caption{}
                \label{fig:fig5c}
        \end{subfigure}
        \caption{Potential energy curves of a) H$_2$, b) CH, and c) C$_2$ diatomics.  The MEAM results are compared to those of REBO, ReaxFF, and the CCSD(2)/aug-cc-pVTZ FP data generated in this work.  The double arrow indicates the coordinate that is being varied.}
\label{fig:fig5}
\end{figure}

The potential energy curves for H$_2$, CH, and C$_2$ diatomics generated by the MEAM, REBO, and ReaxFF potentials are presented in Fig.~\ref{fig:fig5}.  As stated in Section \ref{sec:sec3}, the FP data were generated in this work using CCSD(2) and the aug-cc-pVTZ basis set, and MEAM was fitted to these potential energy curves.  The ReaxFF potential implementation in LAMMPS \cite{Pli1995} gives a non-zero energy value for a single carbon atom, $E=-0.075568$ eV.  The ReaxFF interaction curves for CH and C$_2$ were corrected for this non-zero energy associated with the isolated carbon atom at infinite interatomic distances.  ReaxFF is much ``softer'' at small interatomic distances than MEAM, REBO, and FP.  Both ReaxFF and REBO have a much shorter range than FP, while MEAM agrees with FP.  For the H$_2$ diatomic, the bond distance at the minimum energy and the minimum energy agree well with FP for all three potentials. The MEAM prediction of the minimum energy for the CH diatomic and the ReaxFF prediction of the minimum energy for the C$_2$ diatomic significantly differ from the FP minimum energy, while REBO predictions are much closer to the FP data.

\clearpage

\subsection{Dimer Molecules}

\begin{figure}[ht!]
        \centering
        \begin{subfigure}[b]{0.45\textwidth}
                \centering
                \includegraphics[width=\textwidth]{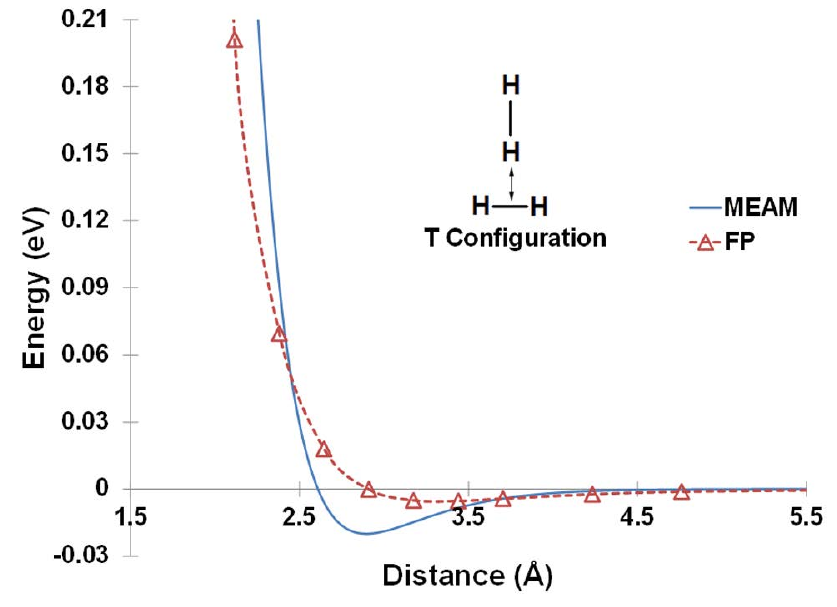}
                \caption{}
                \label{fig:fig6a}
        \end{subfigure}%
\quad
        \begin{subfigure}[b]{0.45\textwidth}
                \centering
                \includegraphics[width=\textwidth]{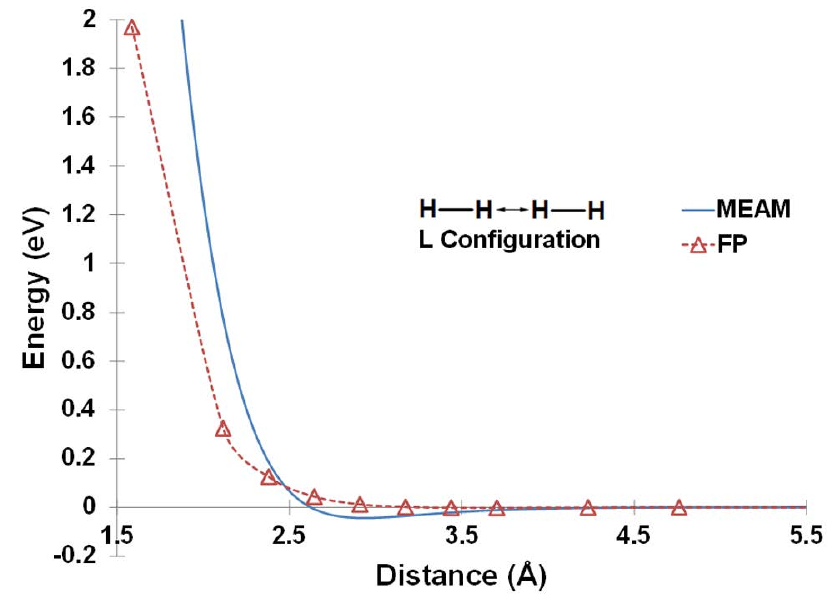}
                \caption{}
                \label{fig:fig6b}
        \end{subfigure}
\\
        \begin{subfigure}[b]{0.45\textwidth}
                \centering
                \includegraphics[width=\textwidth]{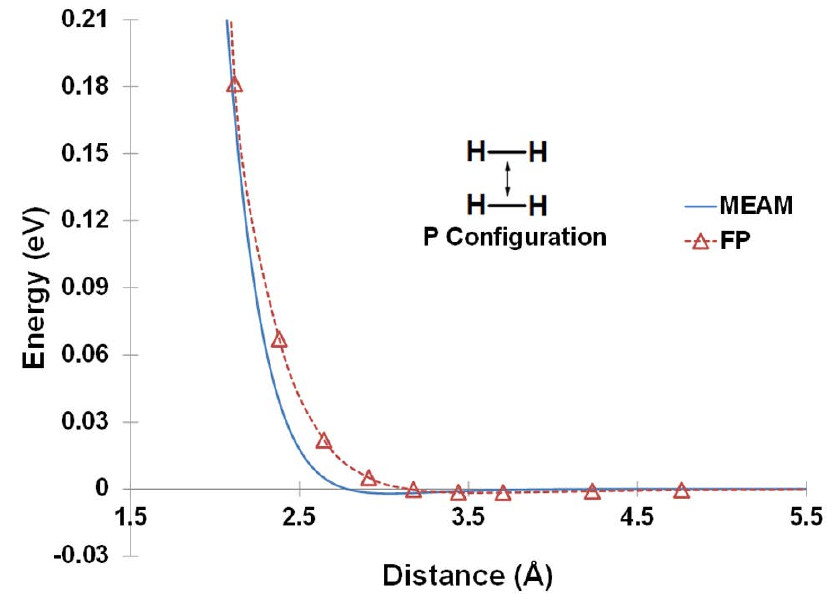}
                \caption{}
                \label{fig:fig6c}
        \end{subfigure}
\quad
        \begin{subfigure}[b]{0.45\textwidth}
                \centering
                \includegraphics[width=\textwidth]{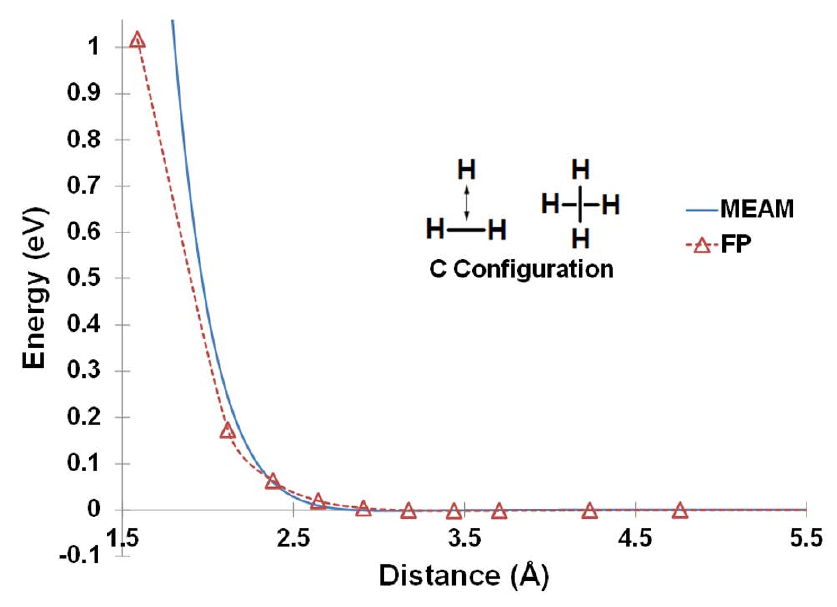}
                \caption{}
                \label{fig:fig6d}
        \end{subfigure}
        \caption{First-principles (FP) \cite{Bur1982} versus MEAM-calculated interaction energy curves for (H$_2$)$_2$ (hydrogen dimer).  The molecular configurations are a) collinear coplanar (T), b) linear (L), c) parallel or rectangular (P), and d) crossed (C) as reported in the work of Burton et~al.~\cite{Bur1982}  The atoms are constrained during energy calculation at each distance increment. The double arrow indicates the coordinate that is being varied.}
\label{fig:fig6}
\end{figure}

\begin{figure}[ht!]
        \centering
        \begin{subfigure}[b]{0.45\textwidth}
                \centering
                \includegraphics[width=\textwidth]{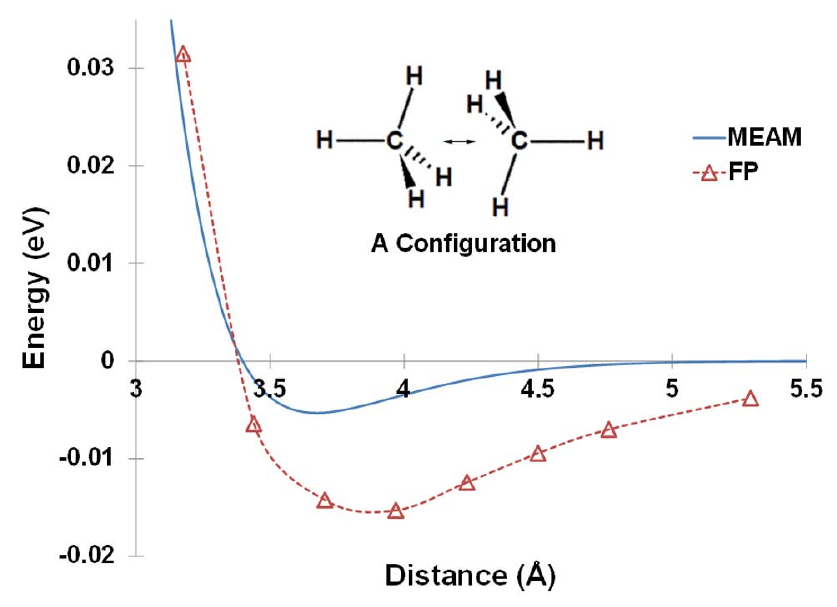}
                \caption{}
                \label{fig:fig7a}
        \end{subfigure}%
\quad
        \begin{subfigure}[b]{0.45\textwidth}
                \centering
                \includegraphics[width=\textwidth]{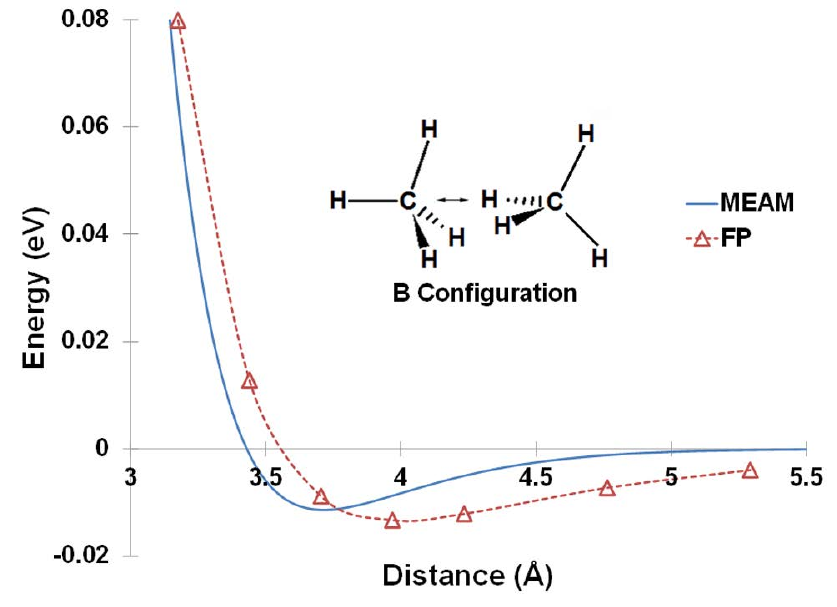}
                \caption{}
                \label{fig:fig7b}
        \end{subfigure}
\\
        \begin{subfigure}[b]{0.45\textwidth}
                \centering
                \includegraphics[width=\textwidth]{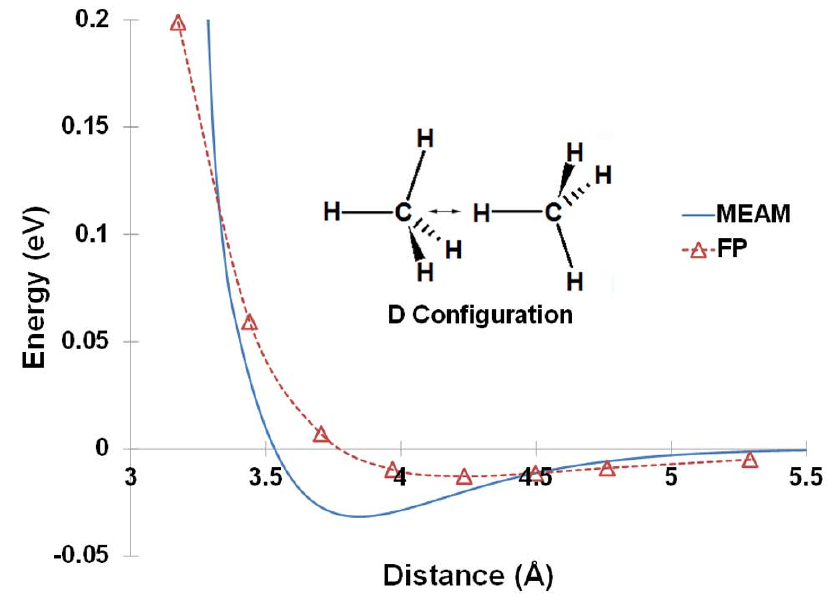}
                \caption{}
                \label{fig:fig7c}
        \end{subfigure}
\quad
        \begin{subfigure}[b]{0.45\textwidth}
                \centering
                \includegraphics[width=\textwidth]{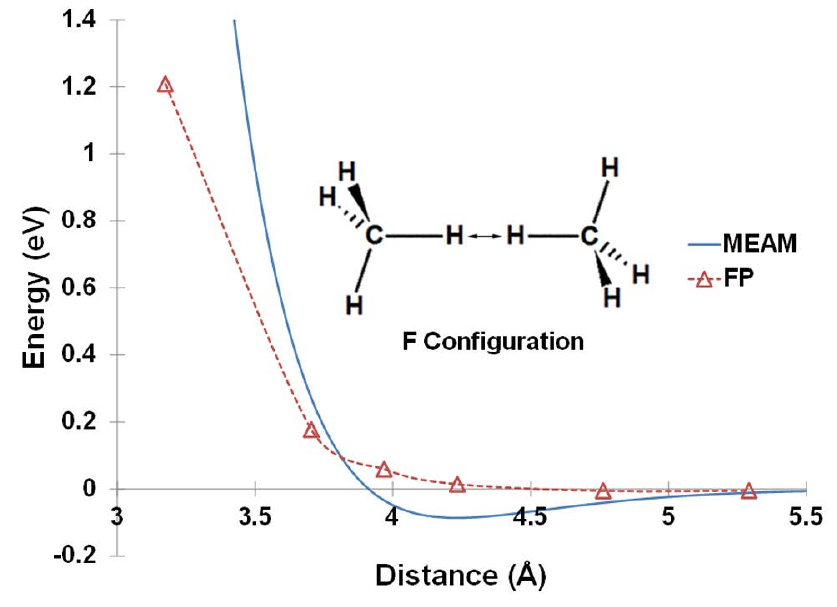}
                \caption{}
                \label{fig:fig7d}
        \end{subfigure}
        \caption{First-principles (FP) \cite{Szc1990} versus MEAM-calculated interaction energy curves for (CH$_4$)$_2$ (methane dimer).  The molecular configurations A, B, D, and F are reported in the work of Szczesniak et~al.~\cite{Szc1990}  The atoms are constrained during energy calculation at each distance increment. The double arrow indicates the coordinate that is being varied.}
\label{fig:fig7}
\end{figure}

\begin{figure}[ht!]
        \centering
        \begin{subfigure}[b]{0.45\textwidth}
                \centering
                \includegraphics[width=\textwidth]{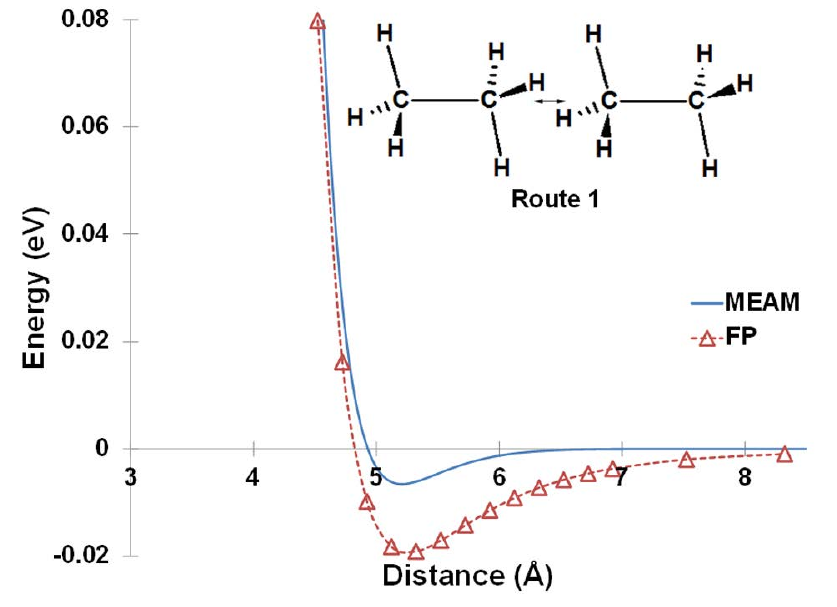}
                \caption{}
                \label{fig:fig8a}
        \end{subfigure}%
\quad
        \begin{subfigure}[b]{0.45\textwidth}
                \centering
                \includegraphics[width=\textwidth]{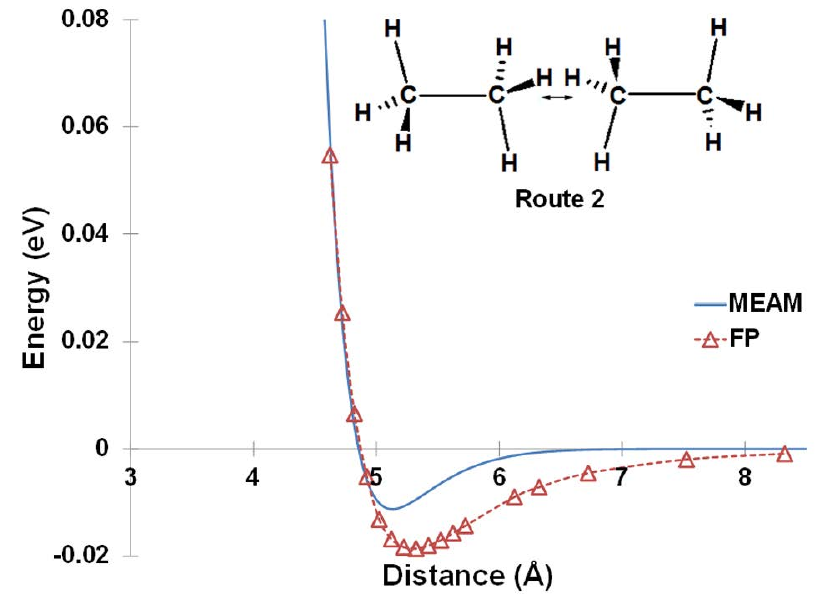}
                \caption{}
                \label{fig:fig8b}
        \end{subfigure}
\\
        \begin{subfigure}[b]{0.45\textwidth}
                \centering
                \includegraphics[width=\textwidth]{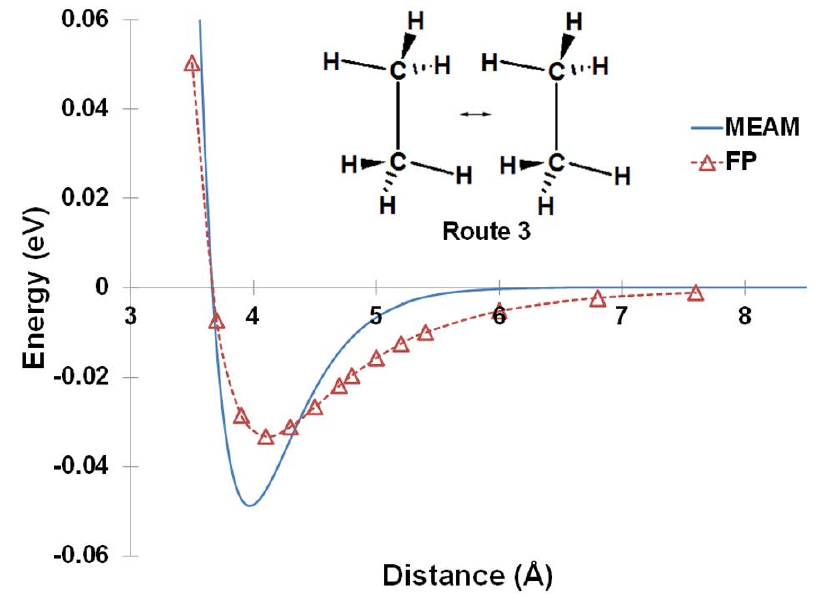}
                \caption{}
                \label{fig:fig8c}
        \end{subfigure}
\quad
        \begin{subfigure}[b]{0.45\textwidth}
                \centering
                \includegraphics[width=\textwidth]{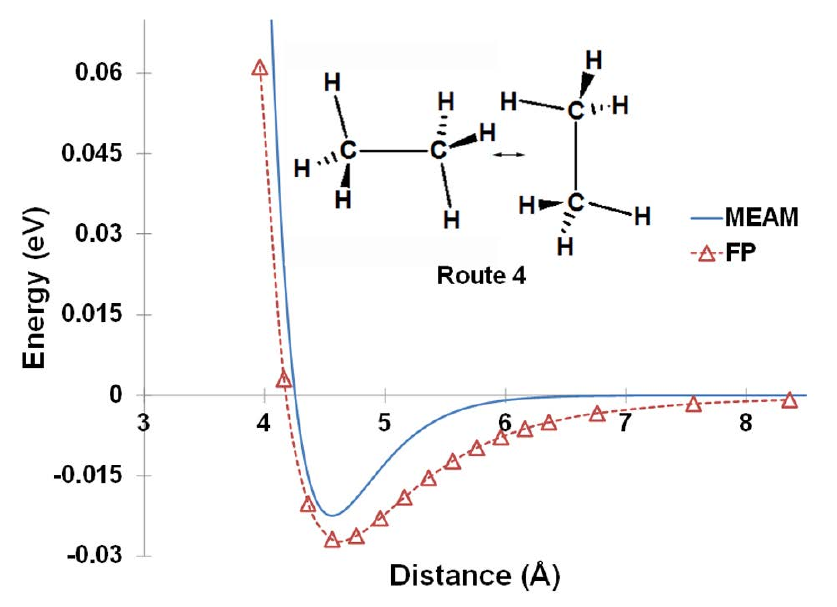}
                \caption{}
                \label{fig:fig8d}
        \end{subfigure}
        \caption{First-principles (FP) \cite{Row2001} versus MEAM-calculated interaction energy curves for (C$_2$H$_6$)$_2$ (ethane dimer).  The molecular configurations (Routes 1-4) are reported in the work of Rowley and Yang. \cite{Row2001}  The atoms are constrained during energy calculation at each distance increment. The double arrow indicates the coordinate that is being varied.}
\label{fig:fig8}
\end{figure}

\begin{figure}[ht!]
        \centering
        \begin{subfigure}[b]{0.45\textwidth}
                \centering
                \includegraphics[width=\textwidth]{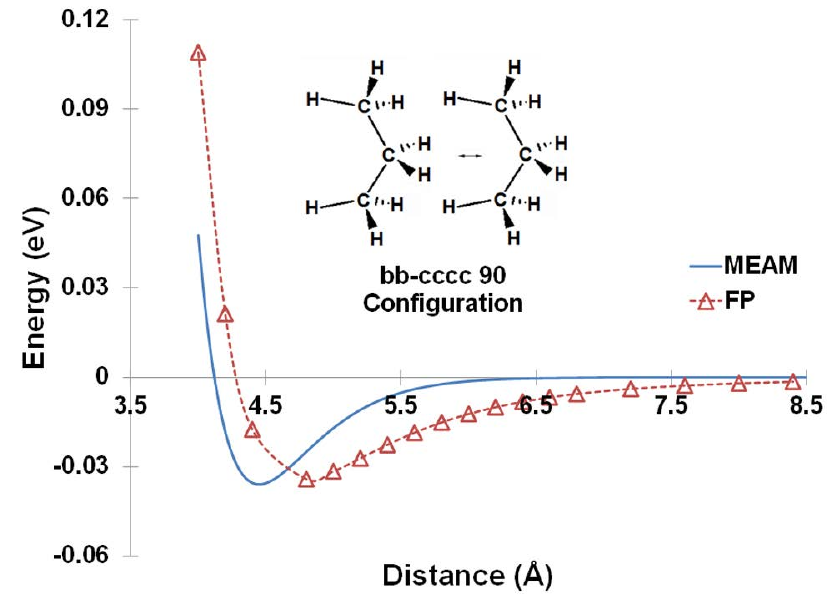}
                \caption{}
                \label{fig:fig9a}
        \end{subfigure}%
\quad
        \begin{subfigure}[b]{0.45\textwidth}
                \centering
                \includegraphics[width=\textwidth]{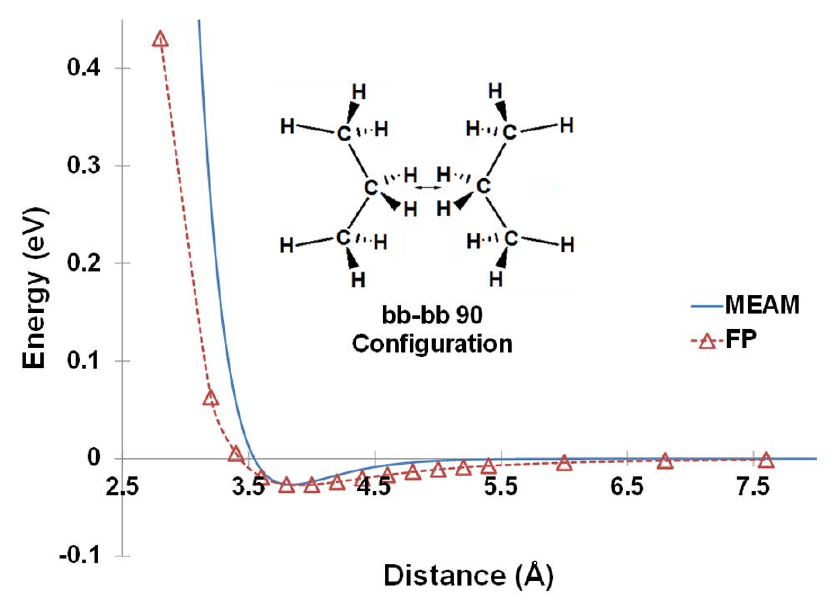}
                \caption{}
                \label{fig:fig9b}
        \end{subfigure}
\quad
        \begin{subfigure}[b]{0.45\textwidth}
                \centering
                \includegraphics[width=\textwidth]{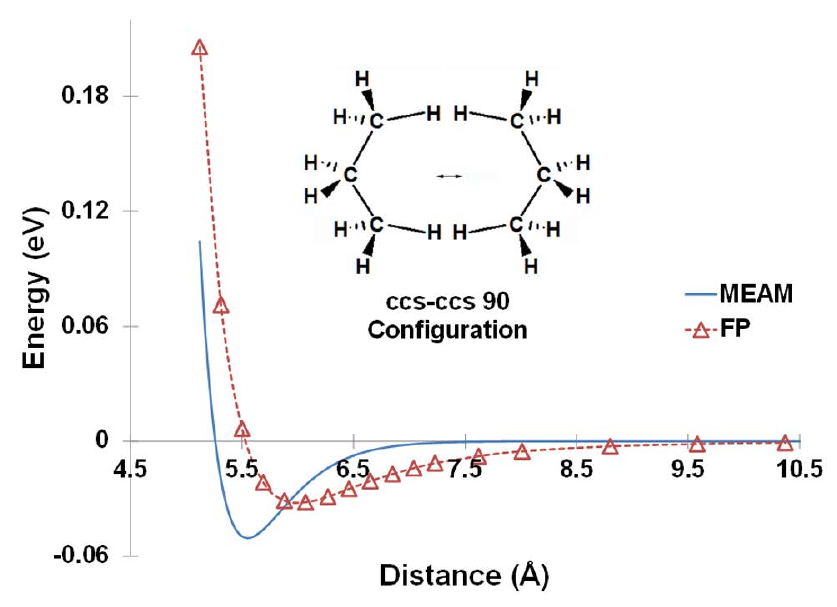}
                \caption{}
                \label{fig:fig9c}
        \end{subfigure}
        \caption{First-principles (FP) \cite{Jal2002} versus MEAM-calculated interaction energy curves for (C$_3$H$_8$)$_2$ (propane dimer).  The molecular configurations (bb-cccc 90, bb-bb 90, and ccs-ccs 90) are reported in the work of Jalkanen et~al.~\cite{Jal2002}  The atoms are constrained during energy calculation at each distance increment. The double arrow indicates the coordinate that is being varied.}
\label{fig:fig9}
\end{figure}

To reproduce the interactions between the molecules, we used the FP interaction energy curves of hydrogen, methane, ethane, and propane dimers in the fitting database (see Section \ref{sec:sec3}).  The MEAM-reproduced interaction curves are plotted for the five dimers in select molecular orientations versus the FP data in Figs.~\ref{fig:fig6}-\ref{fig:fig9}.  All atoms were constrained at each distance increment, and the energy was calculated by subtracting the total energy of the structure at infinite atomic distance from the actual energy at each distance increment.  The FP data for the hydrogen dimer were taken from the work of Burton et~al.~\cite{Bur1982}, where four different molecular configurations, denoted as collinear coplanar (T), linear (L), parallel or rectangular (P), and crossed or elongated tetrahedron (C), were considered.  The configurations are depicted in Fig.~\ref{fig:fig6}.  The FP data for methane and ethane dimers were taken from the work of Szczesniak et~al.~\cite{Szc1990} and Rowley and Yang, \cite{Row2001} respectively.  In the former, four molecular configurations designated in their work as A, B, D, and F were considered ($R_{CH}=1.091$ \si{\angstrom}).  These configurations are depicted in Fig.~\ref{fig:fig7}.  In the latter, only the first four configurations out of 22 reported in the work of Rowley and Yang, designated as routes 1--4, were used for fitting purposes \cite{Row2001} ($R_{CH}=1.102$ \si{\angstrom},  $R_{CC}=1.523$ \si{\angstrom}).  These routes are given as insets in Fig.~\ref{fig:fig8}.  The FP data for propane dimer were taken from the work of Jalkanen et~al.~\cite{Jal2002} with three molecular orientations, designated as bb-cccc90, bb-bb 90, and ccs-ccs 90, used for fitting purposes ($R_{CH}=1.102$ \si{\angstrom}, $R_{CC}=1.529$ \si{\angstrom}).  For details on these orientations and the relevant coding of them, refer to Jalkanen et~al.~\cite{Jal2002}  Note that no additional van der Waals term is added to the MEAM formalism presented above.  The long-range interactions in the MEAM formalism is the subject of future work.

\clearpage

\subsection{Bond Dissociation Reactions and Rotational Barrier}

\begin{figure}[ht!]
        \centering
        \begin{subfigure}[b]{0.6\textwidth}
                \centering
                \includegraphics[width=\textwidth]{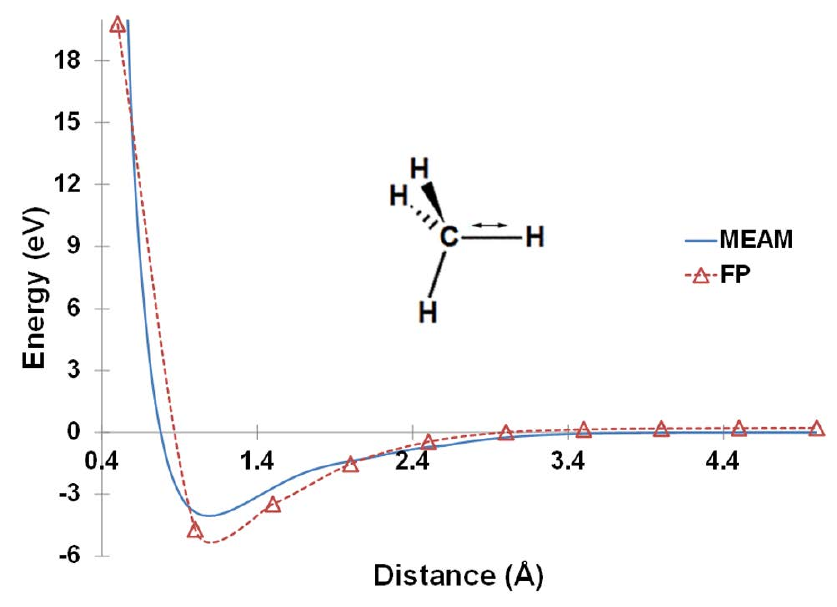}
                \caption{}
                \label{fig:fig10a}
        \end{subfigure}%
\quad
        \begin{subfigure}[b]{0.6\textwidth}
                \centering
                \includegraphics[width=\textwidth]{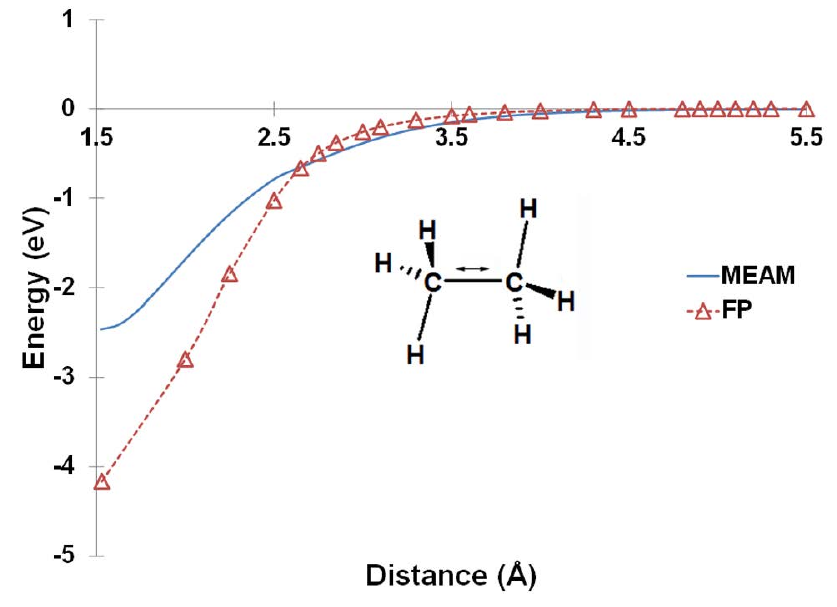}
                \caption{}
                \label{fig:fig10b}
        \end{subfigure}
        \caption{The potential energy curves for the C--H bond dissociation in methane (a) and the C--C bond dissociation in ethane predicted by MEAM versus the FP data.  The FP data for a) were generated in this work, while they were taken from the work of Lorant et~al.~\cite{Lor2001} for b).  The C and H atoms in CH$_4$ and both C atoms in C$_2$H$_6$ were constrained, while the energy was minimized at each distance increment.  The double arrow indicates the coordinate that is being varied.}
\label{fig:fig10}
\end{figure}

\begin{figure}[ht!]
        \centering
                \includegraphics[width=0.75\textwidth]{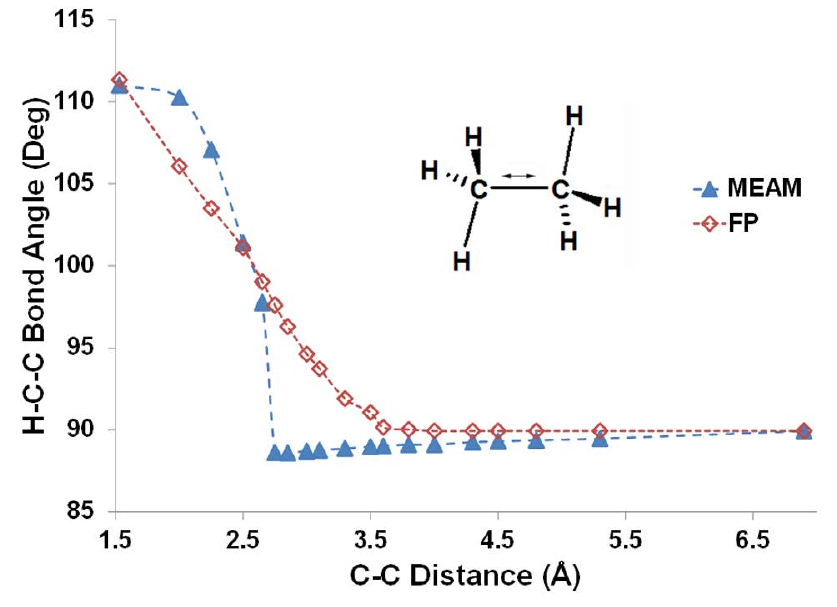}
        \caption{MEAM-reproduced intermediate H--C--C bond angles during the C--C bond dissociation in ethane compared to the FP data reported in the work of Lorant et~al.~\cite{Lor2001}  Both carbon atoms in ethane were constrained, while the energy was minimized at each distance increment.  The double arrow indicates the coordinate that is being varied.}
\label{fig:fig11}
\end{figure}

The MEAM-reproduced C--H and C--C bond dissociation energy curves in methane and ethane and the associated FP data are shown in Figs.~\ref{fig:fig10a} and \ref{fig:fig10b}, respectively.  The FP curve for the C--H dissociation in methane was generated in this work, while the FP data for the C--C bond dissociation in ethane was taken from the work of Lorant et~al.~\cite{Lor2001}  The separating carbon and hydrogen atoms in methane and the two carbon atoms in ethane were constrained, while the structure was minimized at each distance increment. For the FP data of the C--H dissociation, the constrained geometries were minimized with CCSD(T)/aug-cc-pVTZ, and single point energies were calculated with CCSD(2)/aug-cc-pVTZ basis set.  The MEAM-generated data show reasonable agreement with the FP results, especially at longer bond distances.  To further validate the dissociating geometry of the ethane molecule, the intermediate MEAM-calculated $\angle$H--C--C bond angles are compared to the FP data \cite{Lor2001} in Fig.~\ref{fig:fig11}.  The general trend of the FP dissociation curves (Fig.~\ref{fig:fig10}) is qualitatively captured by MEAM, while the angles at intermediate distances deviates some from FP calculations. However, the beginning and end states of the dissociated molecule have the correct $\angle$H--C--C bond angles at similar C--C dissociation distances.  In systems consisting purely of saturated hydrocarbons, C--H and C--C bond dissociations are the only reactions possible.  The heat of reaction data calculated by the MEAM, REBO, and ReaxFF potentials as well as the experimental data for few chemical reactions in saturated hydrocarbons are given in Table \ref{table7}.  In this table, ReaxFF gives the least RMS error (0.20 eV) followed by REBO (0.46 eV) and MEAM (1.23 eV).  The results in Figs.~\ref{fig:fig10}-\ref{fig:fig11} and Table \ref{table7} in combination with data given in Table \ref{table3} give confidence that MEAM is suitable for describing the structures, energetics, and reactions of saturated hydrocarbon systems.

\begin{table}[htbp]
  \centering
  \caption{Heat of reaction ($\Delta{E}_{reaction}$) for selected chemical reactions involving C--H and C--C bond breaking reproduced by the MEAM, REBO, and ReaxFF potentials and compared to the experimental data.}
\begin{threeparttable}[b]    
\begin{tabular}{cC{1.5cm}C{1.5cm}C{1.5cm}C{1.5cm}}
    \hline
    \hline \\ [-2ex]
    \multicolumn{1}{c}{\multirow{2}[4]{*}{Reaction}} & \multicolumn{4}{c}{$\Delta{E}_{reaction}$ (eV)} \\
\cline{2-5} \\ [-2ex]
          & MEAM  & REBO  & ReaxFF & Expt.\tnote{a} \\
    \hline \\ [-2ex]
    \ce{CH4 -> CH3^{.} + H^{.}}  & 3.745 & 4.501 & 4.417 & 4.484 \\
    \ce{C2H6 -> C2H5^{.} + H^{.}} & 3.715 & 3.949 & 4.288 & 4.313 \\
    \ce{C3H8 -> CH3CH^{.}CH3 + H^{.}} & 3.685 & 3.735 & 4.634 & 4.204 \\
    \ce{C2H6 -> 2CH3^{.}} & 2.192 & 3.827 & 3.668 & 3.817 \\
    \ce{C3H8 -> C2H5^{.} + CH3^{.}} & 2.158 & 3.358 & 3.823 & 3.774 \\
    \ce{C4H10 -> 2C2H5^{.}} & 2.120 & 2.888 & 3.898 & 3.752 \\   [0.5ex]
    \hline \\ [-2ex]
    RMS Error\tnote{b} & 1.23 & 0.46 & 0.20  & - \\
    \hline
    \end{tabular}%
        \begin{tablenotes}%
            \item [a] Calculated from the NIST Computational Chemistry Comparison and Benchmark Database. \cite{Nis2011}
            \item [b] Root-Mean-Square Error.
        \end{tablenotes}
\end{threeparttable}
  \label{table7}%
\end{table}%

\begin{figure}[ht!]
        \centering
        \begin{subfigure}[b]{0.6\textwidth}
                \centering
                \includegraphics[width=\textwidth]{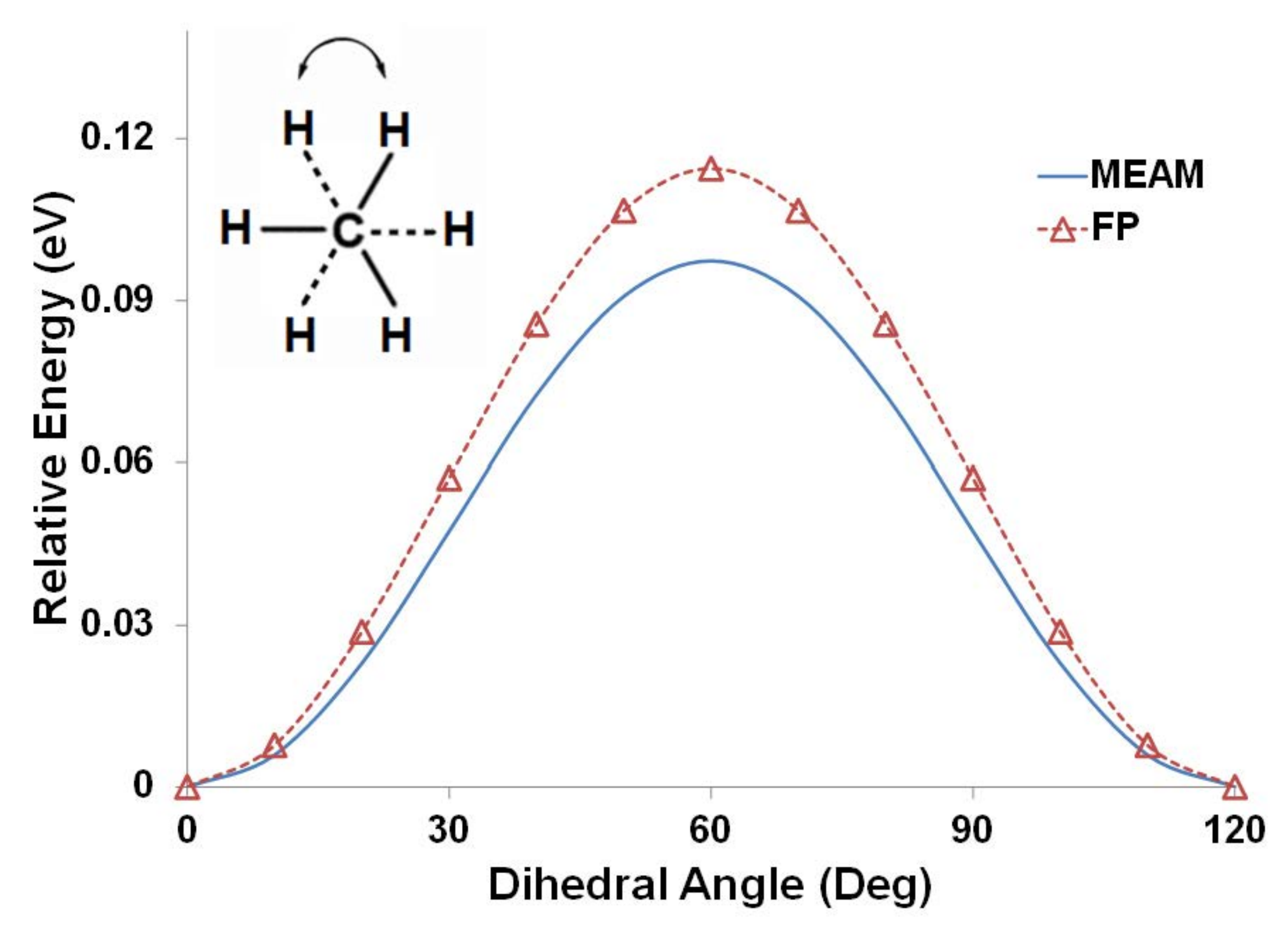}
                \caption{}
                \label{fig:fig12a}
        \end{subfigure}%
\quad
        \begin{subfigure}[b]{0.6\textwidth}
                \centering
                \includegraphics[width=\textwidth]{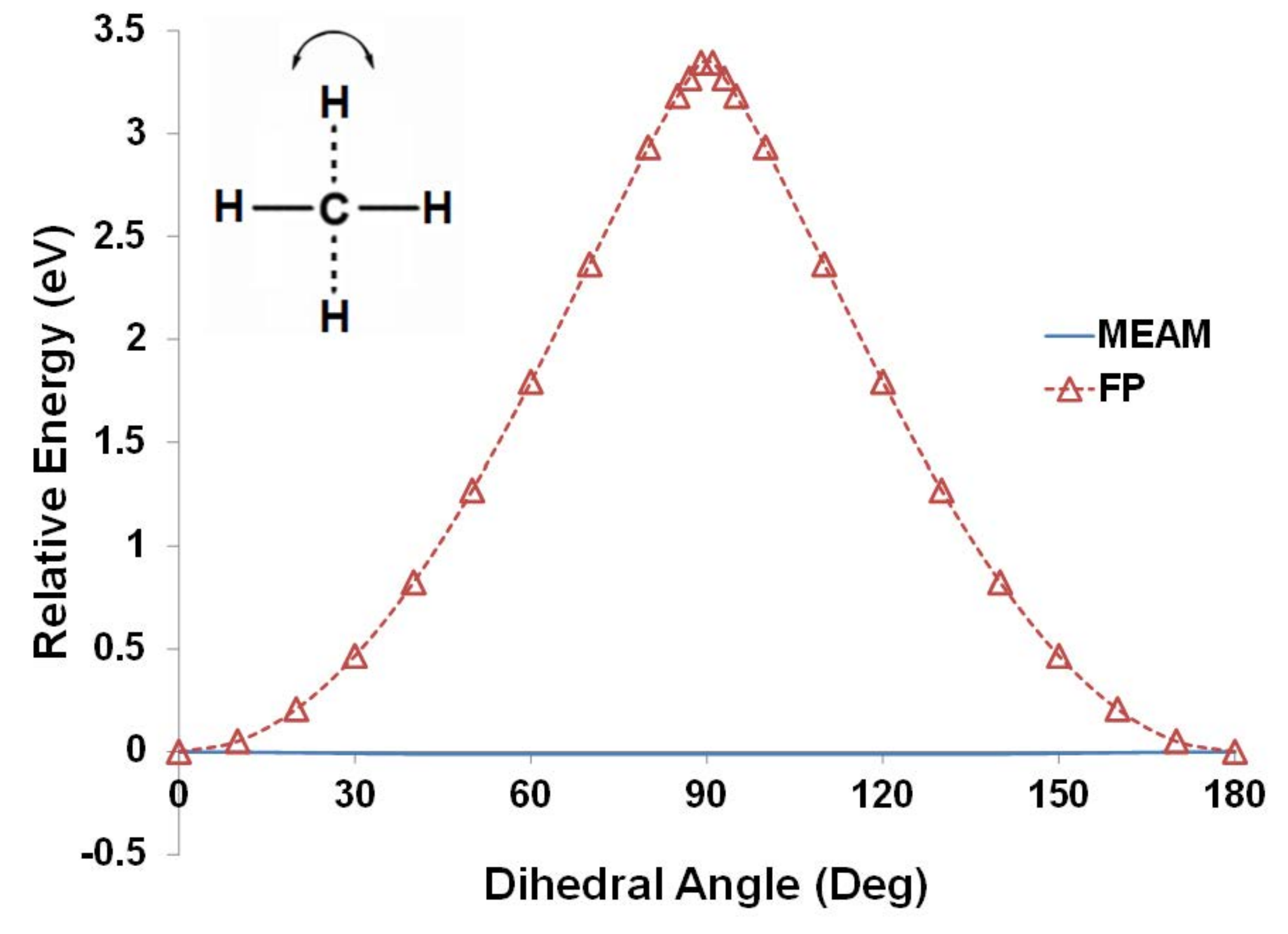}
                \caption{}
                \label{fig:fig12b}
        \end{subfigure}
        \caption{The rotational barrier for a) ethane and b) ethylene calculated by MEAM versus the FP data calculated using  a) CCSD(2)/aug-cc-pVTZ and b) CCSD(T)/aug-cc-pVTZ FP.}
\label{fig:fig12}
\end{figure}

The MEAM-calculated rotational barrier for ethane, ethylene and the associated FP data are given in Fig.~\ref{fig:fig12}.  The FP data calculations are described above.  We have calculated the rotational barrier in ethylene in this work, but obtained a barrier of essentially zero.  Modification of the model to properly reproduce rotation around double bonds is a subject of future research.

\clearpage

\subsection{Molecular Dynamics Simulations}

We further validated our MEAM potential for saturated hydrocarbons with experimental pressure-volume-temperature ($PVT$) data for select alkane systems.  The MEAM-calculated pressure for a high-pressure methane system with the density of 0.5534 g/cc was used for fitting purposes.  A series of MD simulations were run with $NVT$ (constant number of atoms $N$, constant volume $V$, and constant temperature $T$) on a series of 3D periodic methane, ethane, propane, and butane systems with different densities and at different temperatures (Table \ref{table8}).  The total number of atoms in these systems and the cut-off distance (5 \si{\angstrom}) were kept at a minimum due to the large computation times required for running these simulations on the serial DYNAMO \cite{Foi1994} code.  To ensure the reliability of the results with a cut-off distance of 5 \si{\angstrom}, a representative MD simulation was run on a methane system (density of 0.2021 g/cc) with a larger cut-off distance of 9 \si{\angstrom}.  The results of both simulations (using cut-off distances of 5 \si{\angstrom} versus~9 \si{\angstrom}) agree to within 1\%.  Hence, all the rest of simulations were run using the lower 5 \si{\angstrom} cut-off distance.  We built the starting periodic structures in Accelrys\textregistered~Materials Studio\textregistered~software (V5.5) and relaxed them using the COMPASS \cite{Sun1998} force field and Polak-Ribiere conjugate gradient method \cite{Bha2000} for $1000$ iterations.  Then we imported the relaxed periodic structures into DYNAMO and ran the MD simulations on this platform for a total simulation time of 200 ps with a time step of 0.5 fs.  The small time step was chosen to ensure energy conservation in these systems containing the light element hydrogen.  A typical run took from 6--95 hr on a single processor depending on the size and density of the system.  Temperature was controlled by a Nos\'{e}--Hoover thermostat \cite{Nos1984,Hoo1985}.  All systems equilibrated after 50 ps.  The calculated pressures for each time step were time-averaged over the last 150 ps of the simulation, and an average pressure was calculated.  The details of dynamics simulations and the final MEAM-predicted average pressures along with the experimental data are given in Table \ref{table8}.  In spite of typical uncertainties in experimental data and the fact that a small system was used for the MD simulations, MEAM-reproduced average pressures agree well with the experimental data.

\begin{table}[htbp]
  \centering
  \caption{Details on the molecular dynamics simulations of a series of periodic methane, ethane, propane, and butane systems under the $NVT$ ensemble for a total simulation time of 200 ps.  The final MEAM-calculated time-averaged pressure is compared to the experimental data.  The data are reported in the ascending order of density for each molecular species.}
\begin{threeparttable}[b]
    \begin{tabular}{rrrrrrrr}
    \hline
    \hline
    \multicolumn{1}{c}{\multirow{3}[2]{*}{\begin{minipage}{1.75cm}\center{System}\end{minipage}}} & \multicolumn{1}{c}{\multirow{3}[2]{*}{\begin{minipage}{1.5cm}\center{No. of atoms}\end{minipage}}} & \multicolumn{1}{c}{\multirow{3}[2]{*}{\begin{minipage}{3.cm}\center{Cell Size (\si{\angstrom}$^3$)}\end{minipage}}} & \multicolumn{1}{c}{\multirow{3}[2]{*}{\begin{minipage}{1.5cm}\center{Density (g/cc)}\end{minipage}}} & \multicolumn{1}{c}{\multirow{3}[2]{*}{\begin{minipage}{1.5cm}\center{Temp. (K)}\end{minipage}}} & \multicolumn{3}{c}{Pressure} \\
\cline{6-8} \\ [-1.5ex]
     &  & \multicolumn{1}{c}{} & \multicolumn{1}{c}{} & \multicolumn{1}{c}{} & \multicolumn{1}{C{1.5cm}}{Expt.} & \multicolumn{1}{C{1.5cm}}{MEAM} & \multicolumn{1}{C{1.5cm}}{Difference} \\
    \multicolumn{1}{c}{} &  & \multicolumn{1}{c}{} & \multicolumn{1}{c}{} & \multicolumn{1}{c}{} & \multicolumn{1}{c}{(MPa)} & \multicolumn{1}{c}{(MPa)} & \multicolumn{1}{c}{(MPa)} \\
\cline{1-8} \\[-1.5ex]
    \multicolumn{1}{c}{methane-1} & \multicolumn{1}{c}{500} & \multicolumn{1}{c}{47.65x47.65x47.65} & \multicolumn{1}{c}{0.0246} & \multicolumn{1}{c}{400} & \multicolumn{1}{c}{5.005\tnote{a}} & \multicolumn{1}{c}{5.09} & \multicolumn{1}{c}{0.09} \\
    \multicolumn{1}{c}{methane-2} & \multicolumn{1}{c}{500} & \multicolumn{1}{c}{42.71x42.71x42.71} & \multicolumn{1}{c}{0.0342} & \multicolumn{1}{c}{305} & \multicolumn{1}{c}{5.001\tnote{a}} & \multicolumn{1}{c}{4.98} & \multicolumn{1}{c}{-0.02} \\
    \multicolumn{1}{c}{methane-3} & \multicolumn{1}{c}{500} & \multicolumn{1}{c}{28.22x28.22x28.22} & \multicolumn{1}{c}{0.1185} & \multicolumn{1}{c}{298} & \multicolumn{1}{c}{14.994\tnote{a}} & \multicolumn{1}{c}{12.75} & \multicolumn{1}{c}{-2.24} \\
    \multicolumn{1}{c}{methane-4} & \multicolumn{1}{c}{500} & \multicolumn{1}{c}{23.62x23.62x23.62} & \multicolumn{1}{c}{0.2021} & \multicolumn{1}{c}{450} & \multicolumn{1}{c}{59.975\tnote{a}} & \multicolumn{1}{c}{58.65} & \multicolumn{1}{c}{-1.33} \\
    \multicolumn{1}{c}{methane-5} & \multicolumn{1}{c}{500} & \multicolumn{1}{c}{19.17x19.17x19.17} & \multicolumn{1}{c}{0.3782} & \multicolumn{1}{c}{338} & \multicolumn{1}{c}{179.829\tnote{a}} & \multicolumn{1}{c}{162.44} & \multicolumn{1}{c}{-17.39} \\
    \multicolumn{1}{c}{methane-6} & \multicolumn{1}{c}{500} & \multicolumn{1}{c}{18.80x18.80x18.80} & \multicolumn{1}{c}{0.4008} & \multicolumn{1}{c}{298} & \multicolumn{1}{c}{188.059\tnote{a}} & \multicolumn{1}{c}{173.59} & \multicolumn{1}{c}{-14.47} \\
    \multicolumn{1}{c}{methane-7} & \multicolumn{1}{c}{500} & \multicolumn{1}{c}{16.88x16.88x16.88} & \multicolumn{1}{c}{0.5534} & \multicolumn{1}{c}{373} & \multicolumn{1}{c}{1000\tnote{b}} & \multicolumn{1}{c}{1002.41} & \multicolumn{1}{c}{2.41} \\
    \multicolumn{1}{c}{ethane-1} & \multicolumn{1}{c}{800} & \multicolumn{1}{c}{51.25x51.25x51.25} & \multicolumn{1}{c}{0.0371} & \multicolumn{1}{c}{308} & \multicolumn{1}{c}{2.550\tnote{c}} & \multicolumn{1}{c}{1.09} & \multicolumn{1}{c}{-1.46} \\
    \multicolumn{1}{c}{ethane-2} & \multicolumn{1}{c}{800} & \multicolumn{1}{c}{26.36x26.36x26.36} & \multicolumn{1}{c}{0.2726} & \multicolumn{1}{c}{308} & \multicolumn{1}{c}{5.387\tnote{c}} & \multicolumn{1}{c}{5.57} & \multicolumn{1}{c}{0.18} \\
    \multicolumn{1}{c}{ethane-3} & \multicolumn{1}{c}{800} & \multicolumn{1}{c}{21.68x21.68x21.68} & \multicolumn{1}{c}{0.4901} & \multicolumn{1}{c}{260} & \multicolumn{1}{c}{31.294\tnote{c}} & \multicolumn{1}{c}{29.22} & \multicolumn{1}{c}{-2.07} \\
    \multicolumn{1}{c}{propane-1} & \multicolumn{1}{c}{880} & \multicolumn{1}{c}{28.62x28.62x28.62} & \multicolumn{1}{c}{0.2497} & \multicolumn{1}{c}{325} & \multicolumn{1}{c}{26.891\tnote{d}} & \multicolumn{1}{c}{4.28} & \multicolumn{1}{c}{-22.61} \\
    \multicolumn{1}{c}{propane-2} & \multicolumn{1}{c}{880} & \multicolumn{1}{c}{22.40x22.40x22.40} & \multicolumn{1}{c}{0.5212} & \multicolumn{1}{c}{280} & \multicolumn{1}{c}{1.466\tnote{e}} & \multicolumn{1}{c}{4.49} & \multicolumn{1}{c}{3.02} \\
    \multicolumn{1}{c}{$n$-butane} & \multicolumn{1}{c}{700} & \multicolumn{1}{c}{20.23x20.23x20.23} & \multicolumn{1}{c}{0.5827} & \multicolumn{1}{c}{300} & \multicolumn{1}{c}{7.089\tnote{e}} & \multicolumn{1}{c}{-5.10} & \multicolumn{1}{c}{-12.19} \\ [0.5ex]
    \hline 
    \end{tabular}%
        \begin{tablenotes}%
\item [a] From Cristancho et~al.~\cite{Cri2009}
\item [b] From Robertson and Babb~\cite{Rob1969}
\item [c] From Straty and Tsumura~\cite{Str1976}
\item [d] From Straty and Palavra~\cite{Str1984}
\item [e] From Kayukawa et~al.~\cite{Kay2005}
        \end{tablenotes}
\end{threeparttable}
  \label{table8}%
\end{table}%

\clearpage

\section{Concluding Remarks}

We have successfully developed a new semi-empirical many-body potential for saturated hydrocarbons based on the modified embedded-atom method.  The potential parameterization was performed with respect to a large database of atomization energies, bond distances, and bond angles of a homologous series of alkanes and their isomers up to $n$-octane, the potential energy curves of H$_2$, CH, and C$_2$, (H$_2$)$_2$, (CH$_4$)$_2$, (C$_2$H$_6$)$_2$, and (C$_3$H$_8$)$_2$ and the pressure-volume-temperature ($PVT$) relationship of a dense methane system.  The new potential successfully predicts the $PVT$ behavior of representative alkane systems at different densities and temperatures.  Furthermore, MEAM predicts the energetics and geometries of the methane and ethane molecules undergoing a bond-breaking reaction reasonably well.  The significance of this work is in the extension of the classical MEAM formalism for metals and metal hydride, carbide, and nitride systems to saturated hydrocarbons.  This is the first step toward its universality for all atomic and molecular systems.  The main benefit of using this potential versus other potentials for various atomic and molecular dynamics simulation studies is its vast parameter database for metals.  This makes it possible, for example, to study complex polymer-metal systems using the same formalism for both metals and organic molecules.  In addition, MEAM is inherently linear scaling, making possible simulations on very large systems.  Since MEAM is a reactive potential, numerous possible simulation studies of reactive organic/metal systems as well as void and crack formation and growth in polymer systems are envisioned.

\section*{Acknowledgments}

The authors thank the Department of Energy for partially supporting this work under contract DE-FC26-06NT42755.  MAT acknowledges support from the U.S. Army Research Laboratory.

\bibliographystyle{unsrt}

\begin{thebibliography}{10}

\bibitem{Daw1983}
M.S. Daw and M.I. Baskes.
\newblock Semiempirical, quantum mechanical calculation of hydrogen
  embrittlement in metals.
\newblock {\em Physical Review Letters}, 50(17):1285--1288, 1983.

\bibitem{Daw1984}
M.S. Daw and M.I. Baskes.
\newblock Embedded-atom method: Derivation and application to impurities,
  surfaces, and other defects in metals.
\newblock {\em Physical Review B}, 29(12):6443, 1984.

\bibitem{Daw1993}
M.S. Daw, S.M. Foiles, and M.I. Baskes.
\newblock The embedded-atom method:~{A} review of theory and applications.
\newblock {\em Materials Science Reports}, 9(7-8):251--310, 1993.

\bibitem{Bas1987}
M.I.~Baskes.
\newblock Application of the embedded-atom method to covalent materials: A
  semiempirical potential for silicon.
\newblock {\em Physical Review Letters}, 59(23):2666--2669, 1987.

\bibitem{Bas1992}
M.I. Baskes.
\newblock Modified embedded-atom potentials for cubic materials and impurities.
\newblock {\em Physical Review B}, 46(5):2727--2742, 1992.

\bibitem{Bas1989}
M.I. Baskes, J.S. Nelson, and A.F. Wright.
\newblock Semiempirical modified embedded-atom potentials for silicon and
  germanium.
\newblock {\em Physical Review B}, 40(9):6085--6100, 1989.

\bibitem{Lee2000}
B.J. Lee and M.I.~Baskes.
\newblock Second nearest-neighbor modified embedded-atom-method potential.
\newblock {\em Physical Review B}, 62(13):8564, 2000.

\bibitem{Lee2001}
B.J. Lee, M.I. Baskes, H.~Kim, and Y.K. Cho.
\newblock Second nearest-neighbor modified embedded atom method potentials for
  bcc transition metals.
\newblock {\em Physical Review B}, 64(18):184102, 2001.

\bibitem{Lee2003}
B.J. Lee, J.H. Shim, and M.I. Baskes.
\newblock Semiempirical atomic potentials for the fcc metals {Cu, Ag, Au, Ni,
  Pd, Pt, Al, and Pb} based on first and second nearest-neighbor modified
  embedded atom method.
\newblock {\em Physical Review B}, 68(14):144112, 2003.

\bibitem{Bas2007}
M.I.~Baskes, S.G.~Srinivasan, S.M.~Valone, and R.G.~Hoagland.
\newblock Multistate modified embedded atom method.
\newblock {\em Physical Review B}, 75(9):094113, 2007.

\bibitem{Zha2004}
J.M. Zhang, F.~Ma, and K.W. Xu.
\newblock Calculation of the surface energy of fcc metals with modified
  embedded-atom method.
\newblock {\em Applied Surface Science}, 229(1):34--42, 2004.

\bibitem{Zha2003a}
J.M. Zhang, F.~Ma, K.W. Xu, and X.T. Xin.
\newblock Anisotropy analysis of the surface energy of diamond cubic crystals.
\newblock {\em Surface and Interface Analysis}, 35(10):805--809, 2003.

\bibitem{Bas1994}
M.I.~Baskes and R.A.~Johnson.
\newblock Modified embedded atom potentials for hcp metals.
\newblock {\em Modelling and Simulation in Materials Science and Engineering},
  2:147, 1994.

\bibitem{Hu2001}
W.~Hu, B.~Zhang, B.~Huang, F.~Gao, and D.J. Bacon.
\newblock Analytic modified embedded atom potentials for hcp metals.
\newblock {\em Journal of Physics: Condensed Matter}, 13(6):1193, 2001.

\bibitem{Zha2003}
J.M. Zhang, F.~Ma, and K.W. Xu.
\newblock Calculation of the surface energy of bcc metals by using the modified
  embedded-atom method.
\newblock {\em Surface and Interface Analysis}, 35(8):662--666, 2003.

\bibitem{Jel2012}
B.~Jelinek, S.~Groh, M.F.~Horstemeyer, J.~Houze, S.G.~Kim, G.J.~Wagner, A.~Moitra,
  and M.I.~Baskes.
\newblock Modified embedded atom method potential for {Al, Si, Mg, Cu, and Fe}
  alloys.
\newblock {\em Physical Review B}, 85(24):245102, 2012.

\bibitem{Kim2009}
H.K. Kim, W.S. Jung, and B.J. Lee.
\newblock Modified embedded-atom method interatomic potentials for the
  {Fe--Ti--C} and {Fe--Ti--N} ternary systems.
\newblock {\em Acta Materialia}, 57(11):3140--3147, 2009.

\bibitem{Hor2012}
M.F. Horstemeyer.
\newblock {\em Integrated Computational Materials Engineering (ICME) for
  Metals: Using Multiscale Modeling to Invigorate Engineering Design with
  Science}.
\newblock Wiley, 2012.

\bibitem{Lee2010}
B.J. Lee, W.S. Ko, H.K. Kim, and E.H. Kim.
\newblock The modified embedded-atom method interatomic potentials and recent
  progress in atomistic simulations.
\newblock {\em Calphad - Computer Coupling of Phase Diagrams and
  Thermochemistry}, 34(4):510--522, 2010.

\bibitem{Xia2009}
W.~Xiao, M.I.~Baskes, and K.~Cho.
\newblock {MEAM} study of carbon atom interaction with {Ni} nano particle.
\newblock {\em Surface Science}, 603(13):1985--1998, 2009.

\bibitem{Udd2010}
J. Uddin, M.I. Baskes, S.G. Srinivasan, T.R. Cundari, and A.K.
  Wilson.
\newblock Modified embedded atom method study of the mechanical properties of
  carbon nanotube reinforced nickel composites.
\newblock {\em Physical Review B}, 81(10):104103, 2010.

\bibitem{All1989}
N.L. Allinger, Y.H. Yuh, and J.H. Lii.
\newblock Molecular mechanics. {The MM3} force field for hydrocarbons. 1.
\newblock {\em Journal of the American Chemical Society}, 111(23):8551--8566,
  1989.

\bibitem{Lii1989}
J.H. Lii and N.L. Allinger.
\newblock Molecular mechanics. {The MM3} force field for hydrocarbons. 2.
  {V}ibrational frequencies and thermodynamics.
\newblock {\em Journal of the American Chemical Society}, 111(23):8566--8575,
  1989.

\bibitem{Lii1989a}
J.H. Lii and N.L. Allinger.
\newblock Molecular mechanics. {The MM3} force field for hydrocarbons. 3. {The
  van der Waals'} potentials and crystal data for aliphatic and aromatic
  hydrocarbons.
\newblock {\em Journal of the American Chemical Society}, 111(23):8576--8582,
  1989.

\bibitem{All1996}
N.L. Allinger, K.~Chen, and J.H. Lii.
\newblock An improved force field ({MM}4) for saturated hydrocarbons.
\newblock {\em Journal of Computational Chemistry}, 17(5-6):642--668, 1996.

\bibitem{May1990}
S.L. Mayo, B.D. Olafson, and W.A. Goddard.
\newblock {DREIDING}: {A} generic force field for molecular simulations.
\newblock {\em Journal of Physical Chemistry}, 94(26):8897--8909, 1990.

\bibitem{Bre1990}
D.W. Brenner.
\newblock Empirical potential for hydrocarbons for use in simulating the
  chemical vapor deposition of diamond films.
\newblock {\em Physical Review B}, 42(15):9458, 1990.

\bibitem{Bre2002}
D.W. Brenner, O.A. Shenderova, J.A. Harrison, S.J. Stuart, B.~Ni, and S.B.
  Sinnott.
\newblock A second-generation reactive empirical bond order {(REBO)} potential
  energy expression for hydrocarbons.
\newblock {\em Journal of Physics: Condensed Matter}, 14:783, 2002.

\bibitem{Van2001}
A.C.T. Van~Duin, S.~Dasgupta, F.~Lorant, and W.A. Goddard~III.
\newblock {ReaxFF}: a reactive force field for hydrocarbons.
\newblock {\em Journal of Physical Chemistry A}, 105(41):9396--9409, 2001.

\bibitem{Yu2007}
J. Yu, S.B. Sinnott, and S.R. Phillpot.
\newblock Charge optimized many-body potential for the {Si/SiO$_2$} system.
\newblock {\em Physical Review B}, 75(8):085311, 2007.

\bibitem{Shan2010}
T.-R. Shan, B.D. Devine, J.M. Hawkins, A. Asthagiri, S.R. Phillpot, and S.B. Sinnott.
\newblock Second-generation charge-optimized many-body potential for {Si/SiO$_2$} and amorphous silica.
\newblock {\em Physical Review B}, 82(23):235302, 2010.


\bibitem{Sun1998}
H.~Sun.
\newblock {COMPASS}: {A}n ab initio force-field optimized for condensed-phase
  applications overview with details on alkane and benzene compounds.
\newblock {\em Journal of Physical Chemistry B}, 102(38):7338--7364, 1998.

\bibitem{Somers2013}
W. Somers, A. Bogaerts, A.C.T. van Duin, S. Huygh, K.M. Bal, and E.C. Neyts.
\newblock Temperature influence on the reactivity of plasma species on a nickel catalyst surface: An atomic scale study.
\newblock {\em Catalysis Today}, 211(0):131-136, 2013.

\bibitem{Castro2013}
F. Castro-Marcano and A.C.T. van Duin.
\newblock Comparison of thermal and catalytic cracking of 1-heptene from {ReaxFF} reactive molecular dynamics simulations.
\newblock {\em Combustion and Flame}, 160(4):766-775, 2013.

\bibitem{Monti2013}
S. Monti, C. Li, and V. Carravetta.
\newblock Dynamics simulation of monolayer and multilayer adsorption of glycine on {Cu(110)}.
\newblock {\em Journal of Physical Chemistry C}, 117(10):5221-5228, 2013.

\bibitem{Kim2013}
S.-Y. Kim, A.C.T. van Duin, and J.D. Kubicki.
\newblock Molecular dynamics simulations of the interactions between {TiO$_2$} nanoparticles and water with {Na+} and {Cl−}, methanol, and formic acid using a reactive force field.
\newblock {\em Journal of Materials Research}, 28(03):513-520, 2013.

\bibitem{Liang2012}
T. Liang, B. Devine, S.R. Phillpot, and S.B. Sinnott.
\newblock Variable charge reactive potential for hydrocarbons to simulate organic-copper interactions.
\newblock {\em Journal of Physical Chemistry A}, 116(30):7976-7991, 2012.

\bibitem{Liang2013}
T. Liang, Y.K. Shin, Y.-T. Cheng, D.E. Yilmaz, K.G. Vishnu, O. Verners, C. Zou, S.R. Phillpot, S.B. Sinnott, A.C.T. van Duin.
\newblock Reactive potentials for advanced atomistic simulations.
\newblock {\em Annual Review of Materials Research}, 43:109-129, 2013.

\bibitem{Val2006}
S.M.~Valone, M.I.~Baskes, and R.L.~Martin.
\newblock Atomistic model of helium bubbles in gallium-stabilized plutonium
  alloys.
\newblock {\em Physical Review B}, 73(21):214209, 2006.

\bibitem{Bas1999}
M.I.~Baskes.
\newblock Atomistic potentials for the molybdenum-silicon system.
\newblock {\em Materials Science and Engineering A}, 261(1):165--168, 1999.

\bibitem{Ros1984}
J.H. Rose, J.R. Smith, F.~Guinea, and J.~Ferrante.
\newblock Universal features of the equation of state of metals.
\newblock {\em Physical Review B}, 29(6):2963, 1984.

\bibitem{Val2006a}
S.M.~Valone, V. Kapila.
\newblock In {\em Nonequilibrium atomistic polymer simulations under shear and shock loading.}
\newblock AIP Conference Proceedings, p.~425, 2006.

\bibitem{Lid2009}
D.R. Lide.
\newblock {\em CRC handbook of chemistry and physics: a ready-reference book of
  chemical and physical data}.
\newblock CRC Press, 2009.

\bibitem{Nis2011}
{\em NIST Computational Chemistry Comparison and Benchmark Database, NIST
  Standard Reference Database Number 101, Release 15b}.
\newblock http://cccbdb.nist.gov/, August 2011.

\bibitem{Kar2009}
A. Karton, D. Gruzman, and J.M.L. Martin.
\newblock Benchmark thermochemistry of the {C$_n$H$_{2n+2}$} alkane isomers (n
  = 2--8) and performance of {DFT} and composite ab initio methods for
  dispersion-driven isomeric equilibria.
\newblock {\em Journal of Physical Chemistry A}, 113(29):8434--8447, 2009.

\bibitem{Bur1982}
P.G. Burton and U.E. Senff.
\newblock The ({H}$_2$)$_2$ potential surface and the interaction between
  hydrogen molecules at low temperatures.
\newblock {\em Journal of Chemical Physics}, 76(12):6073--6087, 1982.

\bibitem{Szc1990}
M.M.~Szczesniak, G.~Chalansinski, S.M.~Cybulski, and S.~Scheiner.
\newblock Intermolecular potential of the methane dimer and trimer.
\newblock {\em Journal of Chemical Physics}, 93:4243, 1990.

\bibitem{Row2001}
R.L. Rowley, Y.~Yang, and T.A. Pakkanen.
\newblock Determination of an ethane intermolecular potential model for use in
  molecular simulations from ab initio calculations.
\newblock {\em Journal of Chemical Physics}, 114:6058, 2001.

\bibitem{Jal2002}
J.-P. Jalkanen, R. Mahlanen, T.A. Pakkanen, and R.L. Rowley.
\newblock Ab initio potential energy surfaces of the propane dimer.
\newblock {\em Journal of Chemical Physics}, 116:1303, 2002.

\bibitem{Rob1969}
S.L. Robertson and S.E. Babb.
\newblock {PVT} properties of methane and propene to 10 kbar and 200\si{\celsius}.
\newblock {\em Journal of Chemical Physics}, 51(4):1357--1361, 1969.

\bibitem{Gwa2001}
S.R. Gwaltney and M. Head-Gordon.
\newblock A second-order perturbative correction to the coupled-cluster singles
  and doubles method: {CCSD} (2).
\newblock {\em Journal of Chemical Physics}, 115:2014, 2001.

\bibitem{Ken1992}
R.A. Kendall, T.H. Dunning~Jr, and R.J. Harrison.
\newblock Electron affinities of the first row atoms revisited. {S}ystematic
  basis sets and wave functions.
\newblock {\em Journal of Chemical Physics}, 96:6796, 1992.

\bibitem{Sha2006}
Y. Shao, L.F. Molnar, Y. Jung, J. Kussmann, C. Ochsenfeld, S.T. Brown, A.T.B. Gilbert, L.V. Slipchenko,
  S.V. Levchenko, D.P. O'Neill, R.A. DiStasio~Jr, R.C. Lochan, T. Wang, G.J.O. Beran, N.A. Besley, J.M. Herbert,
  C.Y.~Lin, T.V.~Voorhis, S.H.~Chien, A. Sodt, R.P. Steele,
  V.A. Rassolov, P.E. Maslen, P.P. Korambath, R.D. Adamson,
  B. Austin, J. Baker, E.F.C. Byrd, H. Dachsel, R.J.
  Doerksen, A. Dreuw, B.D. Dunietz, A.D. Dutoi, T.R.
  Furlani, S.R. Gwaltney, A. Heyden, S. Hirata, C.-P. Hsu, G.
  Kedziora, R.Z. Khalliulin, P. Klunzinger, A.M. Lee, M.S.
  Lee, W.Z. Liang, I. Lotan, N. Nair, B. Peters, E.I. Proynov,
  P.A. Pieniazek, Y.M.~Rhee, J. Ritchie, E. Rosta,
  C.D.~Sherrill, A.C. Simmonett, J.E. Subotnik, H.L.
  Woodcock~Iii, W. Zhang, A.T. Bell, A.K. Chakraborty, D.M.
  Chipman, F.J. Keil, A. Warshel, W.J. Hehre, H.F.
  Schaefer~Iii, J. Kong, A.I. Krylov, P.M.W. Gill, and M.
  Head-Gordon.
\newblock Advances in methods and algorithms in a modern quantum chemistry
  program package.
\newblock {\em Physical Chemistry Chemical Physics}, 8(27):3172--3191, 2006.

\bibitem{Gwa2003}
S.R. Gwaltney, G.J.O. Beran, and M.~Head-Gordon.
\newblock {\em Partitioning Techniques in Coupled-Cluster Theory}, volume~1,
  pages 433--457.
\newblock Kluwer Academic Publishers, Dordrecht, Netherlands, 2003.

\bibitem{Han2012}
M.D. Hanwell, D.E. Curtis, D.C. Lonie, T.~Vandermeersch, E.~Zurek, and G.R.
  Hutchison.
\newblock Avogadro: {A}n advanced semantic chemical editor, visualization, and
  analysis platform.
\newblock {\em Journal of Cheminformatics}, 4:17, 2012.

\bibitem{Hal1996}
T.A. Halgren.
\newblock Merck molecular force field. {I}. {B}asis, form, scope,
  parameterization, and performance of {MMFF94}.
\newblock {\em Journal of Computational Chemistry}, 17(5-6):490--519, 1996.

\bibitem{Bha2000}
A.~Bhatti.
\newblock {\em Practical Optimization Methods: With Mathematica\textregistered~
  Applications}.
\newblock Springer-Verlag, 2000.

\bibitem{Mat2010}
T.R. Mattsson, J.M.D. Lane, K.R. Cochrane, M.P. Desjarlais, A.P. Thompson, F. Pierce, and G.S. Grest.
\newblock First-principles and classical molecular dynamics simulation of shocked polymers.
\newblock {\em Physical Review B}, 81(5):054103, 2008.

\bibitem{Pli1995}
S.~Plimpton.
\newblock Fast parallel algorithms for short-range molecular dynamics.
\newblock {\em Journal of Computational Physics}, 117(1):1--19, 1995.

\bibitem{Chenoweth2008}
K. Chenoweth, A.C.T. van Duin, W.A. Goddard III.
\newblock ReaxFF reactive force field for molecular dynamics simulations of hydrocarbon oxidation.
\newblock {\em Journal of Physical Chemistry A}, 112(5):1040-1053, 2008.


\bibitem{Foi1994}
S.M. Foiles, M.S. Daw, and M.I. Baskes.
\newblock {DYNAMO} code.
\newblock {\em Sandia National Laboratories}, 1994.

\bibitem{Lor2001}
F. Lorant, F. Behar, W.A. Goddard, and Y. Tang.
\newblock Ab initio investigation of ethane dissociation using generalized
  transition state theory.
\newblock {\em Journal of Physical Chemistry A}, 105(33):7896--7904, 2001.

\bibitem{Nos1984}
S. Nos\'{e}.
\newblock A unified formulation of the constant temperature molecular dynamics
  methods.
\newblock {\em Journal of Chemical Physics}, 81:511, 1984.

\bibitem{Hoo1985}
W.G. Hoover.
\newblock Canonical dynamics: Equilibrium phase-space distributions.
\newblock {\em Physical Review A}, 31(3):1695, 1985.

\bibitem{Cri2009}
D.E. Cristancho, I.D. Mantilla, S.~Ejaz, K.R. Hall, M.~Atilhan, and G.A.
  Iglesia-Silva.
\newblock Accurate {P}$\rho${T} data for methane from (300 to 450) {K} up to
  180 {MPa}.
\newblock {\em Journal of Chemical \& Engineering Data}, 55(2):826--829, 2009.

\bibitem{Str1976}
G.C.~Straty and R.~Tsumura.
\newblock {PVT} and vapor pressure measurements on ethane.
\newblock {\em J. Res. Natl. Bur. Stand.(US) A}, 80:35--39, 1976.

\bibitem{Str1984}
G.C.~Straty and A.M.F. Palavra.
\newblock Automated high temperature {PVT} apparatus with data for propane.
\newblock {\em Journal of Research of the National Bureau of Standards}, 89(5):375--383, 1984.

\bibitem{Kay2005}
Y.~Kayukawa, M.~Hasumoto, Y.~Kano, and K.~Watanabe.
\newblock Liquid-phase thermodynamic properties for propane (1), n-butane (2),
  and isobutane (3).
\newblock {\em Journal of Chemical \& Engineering Data}, 50(2):556--564, 2005.


\end{thebibliography}

\end {document}